\documentclass[a4paper,10pt]{revtex4}
\usepackage{epsf,amssymb,amsmath,latexsym,theorem}
\usepackage{graphicx,verbatim,color}

\usepackage[T1]{fontenc}
\usepackage[latin1]{inputenc}

\renewcommand{\theequation}{\arabic{section}.\arabic{equation}}
\def \slas{\kern -6.2pt /}
\def \sla{\kern -5.4pt /}
\def \sl{\kern -4.0pt /}
\def \Cslas{\kern -6.8pt /}
\def \Dslas{\kern -6.8pt /}
\def \slass{\kern -7.4pt /}

\def \ii{{\mathrm{i}}}
\def \d{{\mathrm{d}}}

\def \lcdi{\tilde{\di}}
\def \lcx{\tilde{x}}
\def \pd{\partial}

\def \qb{\bar q}      
\def \Tr{\text{Tr}}  
\def \di{\mathbf{d}} 
\def \ix{\mathbf{x}} 
\def \D{{\cal D}}      
\def \f12{\hbox{\large$\frac{1}{2}$}}

\begin{document}
\title{Heavy Meson Distribution Amplitudes of Definite Geometric Twist\\
with Contribution of 3-Particle Distribution Amplitudes}
\author{B.~Geyer$^1$} 
\email{geyer@itp.uni-leipzig.de}
\author{O.~Witzel$^2$}
\email{witzel@physik.hu-berlin.de}
\affiliation{$^1$Universit\"at Leipzig, Institut f\"ur Theoretische Physik, 
Augustusplatz 10, 04109 Leipzig, Germany\\
$^2$Humboldt Universit\"at zu Berlin, 
Institut f\"ur Physik, Newtonstraße 15, 12489 Berlin, Germany}

\date{\today}

\begin{abstract}
Under the constraints of HQET the equations of motion of heavy meson distribution amplitudes of definite geometric twist, 
using the knowledge of their off-cone structure, are reformulated as a set of algebraic equations. Together with equations due to various Dirac structures various relations between the (sets of) independent two- and three particle distribution amplitudes of definite geometric twist are derived and presented using both the notion of (double) Mellin moments and re-summed non-local distribution amplitudes. Resolving these relations for the independent two-particle moments in terms of three-particle double moments we confirmed the representation of $\Phi_{\pm|n}$ by Kawamura et al.~(Phys.~Lett.~B \textbf{523} (2001) 111).
\end{abstract}

\maketitle

\vspace*{-0.5cm}
\setcounter{equation}{0}
\section{Introduction}
\label{EoMintro}

The intention of the present work is to extend, within the framework of Heavy Quark 
Effective Theory (HQET) \cite{Isgur,HQET,FGGW,HQETRev}, our understanding on heavy 
meson wave functions or, equivalently, light-cone (LC) distribution amplitudes (DA).
In a previous work \cite{Bmeson} we introduced already two- and three particle DAs 
of definite {\em geometric twist} for pseucoscalar heavy mesons and related them to 
the usual LCDAs (of dynamical twist) \cite{Grozin1996,Beneke2000,BF01,Kodaira2001}. 
In addition, we derived various relations connecting separately the two- and three-particle 
LCDAs. Now we like to derive additional relations connecting two- and three-particle 
DAs mutually. Such relations, based on exact operator identities due to the 
quark equations of motion (EoM) and equalities between the Dirac structures of the DAs, 
have been considered first for light meson DAs, see e.g.~Refs.~\cite{Braun1990,Ball1998,Ball1999}, 
and later on, taking into account the simplifications due to HQET constraints, for heavy meson DAs, 
see e.g.~Refs.~\cite{BF01,Kodaira2001,Kodaira2003,HWZ05,HQW2006}. These relations, as long as the DAs are not 
explicitly known off the light-cone, have to be solved as differential equations for 
the heavy meson LCDAs. But now, since from recent work \cite{Geyer2001,Eilers2003,Joerg} 
the decomposition of various non-local QCD tensor operators into operators of geometric 
twist is explicitly known also off-cone all necessary differentiations of the DAs 
can be performed directly and, therefore, we need not to solve any differential equation to 
get the wanted relations.

To be more specific, the two- and three-particle DAs are related to the vacuum-to-meson 
matrix elements of generic bi- and trilocal operators which, for the present consideration, 
are given by 
\begin{align}
\langle 0|\qb(x)U(x,0)\Gamma h_v(0)|B(v) \rangle 
\qquad \mathrm{and}\qquad
\langle 0|\qb(x)U(x,\vartheta x) F_{\mu\nu}(\vartheta x) x^\nu 
U(\vartheta x,0)\Gamma h_v(0)|B(v)\rangle\,,
\label{MEoff}
\end{align}
respectively,
where $h_v(x)$ and $\qb(x)$ are the heavy quark and light antiquark field, respectively, 
$|B(v) \rangle$ is the (pseudoscalar) $B$-meson state of (fixed) momentum $P = Mv$, 
$\Gamma=\{1, \gamma_\alpha, \ii\sigma_{\alpha\beta},\gamma_5,\gamma_5 \gamma_\alpha,\ii\gamma_5\sigma_{\alpha\beta}\}$ is a generic Dirac matrix and 
$U(\kappa_1 x, \kappa_2 x) = {\cal P} \exp \big\{ - \ii g 
\int_{\kappa_2}^{\kappa_1}\!\d\tau x^\mu A_\mu (\tau x)\big\}$ is the usual path ordered 
phase factor ensuring manifest gauge invariance ($g$ is the strong coupling parameter 
and $1 \geq \vartheta \geq 0 $ some parameter). 
Let us remember that in Fock-Schwinger gauge $(xA=0)$, where the phase factor equals unity,
the gauge potential $A_\mu(x)$ is related to the field strength $F_{\mu\nu}$ according to
$
A_\mu(x)= \int_0^1\!\!\d \vartheta\; \vartheta x^\alpha F_{\alpha\mu}(\vartheta x).
$

In accordance with the definition of usual meson LCDAs \cite{Chernyak1984} 
but, additionally, respecting HQET constraints the $B$-meson LCDAs arise by 
parametrizing the off-cone matrix elements (\ref{MEoff}), e.g.,
\begin{align} 
\label{me}
\langle 0| \bar q(\lcx)\, \Gamma\, h_v(0) | B(v) \rangle 
&= {\mathcal K}^a_\Gamma(v,x) \int_{0}^1 \d u\; \varphi_a(u)\; e^{-\ii\,u \,(P\lcx)},
\\
\label{met}
\langle 0|\qb(\lcx) F_{\mu\nu}(\vartheta \lcx)\lcx^\nu \Gamma h_v(0)|B(v)\rangle
&= {\mathcal K}^a_{\Gamma\mu}(v,x) \int_{0}^1 
\!\!\D \underline{u}\;\Upsilon_a(u_1, u_2)\; e^{-\ii(u_1+\vartheta u_2)(P\lcx)}, 
\end{align}
where $\int_0^1 \!\!\D \underline{u}=\!\int_0^1 \!\!\d u_1\int_0^1 \!\!\d u_2$ and
$ \lcx = x - v\big((vx)- \sqrt{(vx)^2-x^2v^2}\big),~  \lcx^2 =0, $ 
defines some light-ray being related to $x$ by a fixed non-null subsidiary four-vector 
which may be identified with the $B$-meson's velocity. 
As indicated, the matrix elements (\ref{me}) and (\ref{met}) are 
represented by the Fourier transform of the LCDAs w.r.t.~the variable $\lcx P$
and additionally parametrized by a kinematic factor 
${\cal K}^{[s]a}_{\{\sigma\}}(v,\lcx)$ 
depending on the tensor structure $\{\sigma\}$ of the nonlocal LC operator. 
These factors are the {\em basic kinematical structures} (of scale dimension $s$ w.r.t.~$x\partial$) of the matrix elements which, in principle, can be read off 
from their parametrization w.r.t.~leading LCDAs of \emph{dynamical} twist since 
at leading order geometric and dynamical twist coincide by construction.
Their explicit form has been introduced in Ref.~\cite{Bmeson}.
The integration ranges result from the fact that, in the framework of
non-local LC expansion \cite{Anikin1978,Zavialov}, these matrix elements are shown to 
be entire analytic functions in the variable $\lcx P$ \cite{Geyer1994} whose 
support is restricted to  $[-1,1]$. Additionally, 
due to the (anti)symmetry of the relevant QCD operators 
the integration range can be restricted to $0 \le u_i \le 1, i=1,2$. 

Conventionally, LCDAs are characterized by its {\em dynamical twist} which, 
roughly speaking, counts powers of $M/Q$ for the various terms in the kinematic 
decomposition of the matrix elements of non-local QCD operators \cite{Jaffe1992}. 
Here, instead we use the original, group theoretically founded definition of 
{\em geometric twist}, $\tau = $ (scale) dimension $d~ - $ (Lorentz) spin $j$, which has  
been introduced for local QCD operators in Ref.~\cite{Gross1971} and generalized 
to non-local tensor operators {\em on the light-cone} in 
Refs.~\cite{Geyer1999,Geyer2000b,Lazar2002}. The decomposition of the non-local 
tensor operators into operators of definite geometric twist leads to corresponding 
decompositions of the LCDAs \cite{Geyer2001,Geyer2000}.

With the aim of applying the EoM we need the twist decomposition of local as well
as non-local QCD tensor operators also {\em off the light-cone} which has been studied 
in Refs.~\cite{Eilers2003,Joerg}.
The decomposition of non-local operators ${\cal O}_{\{\sigma\}}$ with given tensor 
structure $\{\sigma\}$ into an (infinite) sum of non-local tensor operators of definite 
twist $\tau$ formally reads 
\begin{align}
{\cal O}_{\{\sigma\}} 
&= \sum_{\tau} {\cal O}^{(\tau)}_{\{\sigma\}}
\qquad {\rm with} \qquad
{\cal O}^{(\tau)}_{\{\sigma\}} = 
{\cal P}_{\{\sigma\}}^{(\tau)\,\{\sigma'\}} {\cal O}_{\{\sigma'\}}
\qquad \mathrm{and} \qquad
\sum_{\tau} {\cal P}_{\{\sigma\}}^{(\tau)\,\{\sigma'\}}
= \delta_{\{\sigma\}}^{\{\sigma'\}},
\end{align}
with the projection operators ${\cal P}_{\{\sigma\}}^{(\tau)\,\{\sigma'\}}(x,\di)$
explicitly given in Appendix~\ref{Append1} up to second order tensors 
($\di$ is the off-cone generalization of the inner derivative on the light-cone).

Considering bilocal off-cone operators, the corresponding meson DAs of definite geometric 
twist $\tau$, generically denoted by $\varphi^{(\tau)}_a(u)$, are introduced 
according to Ref.~\cite{Geyer2001} (cf.~also~Refs.~\cite{Geyer2000b,Lazar2000})
\begin{align}
\langle 0|{\cal O}_{\{\sigma\}}(x, 0)|B(v) \rangle
= \sum_{\tau} {\cal P}_{\{\sigma\}}^{(\tau)\,\{\sigma'\}}(x,\di)\,
{\cal K}^{[s]a}_{\{\sigma'\}}(v,x)
\int_0^1\!\! \d u\;e^{-\ii u (Px)} \varphi^{(\tau)}_a(u).
\label{BNL}
\end{align}
In the explicit computations, instead of taking the DAs, 
their (double) Mellin moments will be taken
\begin{alignat}{2}
\varphi_{a|n}&=\int_0^1\!\!\d u\; u^n \;\varphi_{a}(u)\,,
\label{MellinMoments}
\\
\Upsilon_{a|n,m} &=
\int_0^1\!\!\D \underline{u}\; u_1^{n-m} u_2^m \;\Upsilon_{a}(u_1,u_2) 
\qquad
\Longrightarrow 
\qquad
\Upsilon_{a|n}(\vartheta) = \sum_{m=0}^n \binom{n}{m}\; \vartheta^m \;\Upsilon_{a|n,m}\,,
\label{DoubleMoments}
\end{alignat}
respectively.
In terms of Mellin moments $\varphi^{(\tau)}_{a|n}$ 
the two-particle matrix elements decompose as follows,
\begin{align}
\langle 0|{\cal O}_{\{\sigma\}}(x, 0)|B(v) \rangle
&= \sum_{\tau} \sum_{n=0}^\infty
{\cal P}_{\{\sigma\}|n+s}^{(\tau)\,\{\sigma'\}}(x,\di)\,
{\cal K}^{[s]a}_{\{\sigma'\}}(v,x)\,
\frac{(-\ii P x)^n}{n!}\; \varphi^{(\tau)}_{a|n},
\label{BL}
\end{align}
where ${\cal P}_{\{\sigma\}|n+s}^{(\tau)\,\{\sigma'\}}(x,\di)$ are the 
corresponding {\em local} off-cone projection operators (Appendix~\ref{Append1}).    
In case of three-particle matrix elements 
$\langle 0|{\cal O}_{\{\sigma\}}(x,\vartheta x, 0)|B(v) \rangle$
the moments $\varphi^{(\tau)}_{a|n}$ have to be replaced simply by
$\Upsilon^{(\tau)}_{a | n}(\vartheta)$. Luckily, they are not required off the 
light-cone.

An essential property of the geometric twist decomposition is, that it is uniquely determined by 
corresponding irreducible tensor representations of the Lorentz group \cite{Geyer1999,Geyer2000b}. 
Hence this method works for 
any type of matrix elements of tensor operators creating a power series in $x^2$ only and is independent, 
whether the matrix elements are defined by time-ordered field products or otherwise. Any logarithmic 
dependence of the matrix elements on $x^2$, either being determined by the light-cone expansion of 
propagators as in \cite{LMS07} or, in terms of $Q^2$, through renormalization by using renormalization  
group equations \cite{Zavialov,Geyer1994} will be implicitly contained in the wave functions resp. DAs 
$\varphi_a(u)$ and $\Upsilon_a(u_1,u_2)$ as already pointed out in Ref.~\cite{BH94}.

The paper is organized as follows. In Sect.~\ref{EoMoprel} we review the 
operator relations resulting from the quark equations of motion, show
the structure of the relevant matrix elements in the trace formalism and 
introduce some operator relations resulting from relations between Dirac structures. 
In Sect.~\ref{EoMdisc} we present the most general off-cone structure of the bilocal axial
vector matrix element in terms of independent (geometric) twist DAs, determine the 
(single) derivation of the required two-particle off-cone DAs, determine the necessary 
three-particle DAs of definite geometric twist and relate them to the conventional ones,
and write the independent structure elements in terms of geometric twist DAs. 
In Sect.~\ref{EoMrelations} all independent relations between two- and three-particle Mellin 
moments are derived. In Sect.~\ref{EoMnonlocalrelations} the corresponding nonlocal relations
between the two- and three-particle DAs are determined. In Sec.~\ref{EoMKodaira} the 
connection with the work of Kawamura et al.~\cite{Kodaira2001} is carried out and 
their result concerning Mellin moments $\Phi_{\pm|n}$ of usual DAs is re-derived; some remarks
concerning the transverse momentum dependence of these DAs are made.
In Sec.~\ref{EoMconcl} some conclusions will be drawn which might be
of phenomenological relevance for $B$-physics. In Appendix A we review the
explicit twist decompositions of relevant (non)local tensor operators.


\setcounter{equation}{0}
\section{Equations of motion and operator relations}
\label{EoMoprel}


The relations we are interested in are obtained by applying the partial derivative 
$\pd_\mu$ to the bilocal operator $\qb(x)\Gamma U(x,0) h_v(0)$ and  
$\qb(x)\gamma_\mu\Gamma U(x,0) h_v(0)$, respectively, off the light-cone. Therefrom,
the following well-known exact operator identities can be derived, 
cf.~Ref.~\cite{Ball1999b},
\begin{align}
\hspace{-.2cm}
\partial_\mu\, \qb(x)\gamma^\mu \Gamma h_v(0) 
&=\qb(x)\! \stackrel{\leftarrow}{D}\!\!\slas\,\Gamma h_v(0)
+\ii \int_0^1\!\!\d\vartheta\; \vartheta \,
\qb(x) F_{\mu\nu}(\vartheta x) x^\nu \gamma^\mu \Gamma h_v(0),
\label{bt1}\\
\hspace{-.2cm}
v^\mu \partial_\mu \,\qb(x)\Gamma h_v(0) 
&= -v^\mu \,\qb(x)\Gamma D_\mu h_v(0)
+\ii\! \int_0^1\!\!\d\vartheta\,(\vartheta-1)\,
\qb(x) F_{\mu\nu}(\vartheta x) x^\nu v^\mu \Gamma h_v(0) 
+ v^\mu \delta_\mu^T \{ \qb(x)\Gamma h_v(0)\},
\label{bt2}
\end{align}
where $D_\mu = \pd_\mu - \ii A_\mu$ and $\delta_\mu^T\{\cdot\}$ is the `total' derivative 
defined by
\begin{align}
 \delta_\mu^T \{\qb(x)\Gamma h_v(0)\} 
 \equiv 
 \frac{\partial}{\partial y^\mu}\qb(x+y)\Gamma h_v(y)\Big|_{y=0}.
\end{align}

On the one hand, because of the well-known constraints of HQET, these equations 
simplify and, on the other hand, there exist various relations between them for 
different $\Gamma$-structures. Namely, due to the EoM for the light (massless) 
 and the heavy quark,
\begin{align}
\qb(x) \stackrel{\leftarrow}{D}\!\!\slas\, =0
\qquad \mathrm{and} \qquad
(vD)h_v=0\,, 
\end{align}
the first terms on the RHS of Eqs.~(\ref{bt1}) and (\ref{bt2}) 
vanish.  Furthermore, using the heavy quark on-shell constraint, 
\begin{align}
v\sla \,h_v = h_v\,,
\label{onshell}
\end{align}
and the well-known identities for the $\gamma$-matrices, especially
\begin{align}
 \gamma^\mu \gamma^\alpha \gamma^\beta
 &= \left( g^{\mu\alpha} g^{\beta\nu} - g^{\mu\beta} g^{\alpha\nu} 
 + g^{\alpha\beta} g^{\mu\nu} \right) \gamma_\nu
 + \ii \epsilon^{\mu\alpha\beta\nu} \gamma_5 \gamma_\nu\,,
\label{Chisholm}
\\
\gamma_\alpha \gamma_\beta
&= g_{\alpha\beta} - \ii \sigma_{\alpha\beta}\,,\qquad
\sigma_{\alpha\beta} = (\ii/2) [\gamma_\alpha, \gamma_\beta]\,,
\label{Sigma}
\end{align}
the pseudoscalar ($\gamma_5$) and skew tensor ($\ii\gamma_5 \sigma_{\alpha\beta}$) structures can be related to the axial vector structure ($\gamma_5\gamma_\alpha$). Some of these relations have been already derived for the geometric twist LCDAs in our previous work \cite{Bmeson}. 
\smallskip

\noindent
{\it (1) Representation of 2- and 3-particle DAs using the trace formalism}\\
Before proving in general that the knowledge of the axial vector structure is sufficient let us show this by using the definitions of the 2- and 3-particle LCDAs in the trace formalism (see e.g.~\cite{Grozin1996,BF01} as well as \cite{Kodaira2001}).

First of all we consider the three vacuum-to-meson matrix elements containing the bilocal quark-antiquark operator for $\Gamma = \gamma_5, \gamma_5\gamma_\alpha, \gamma_5\ii\sigma_{\alpha\beta}$, usually being parametrized by the help of the DAs $\Phi_\pm = \Phi_\pm(vx, x^2)$, together with their relevant derivatives:
\begin{align}
\label{Kodaira_bi}
\langle 0| \qb(\lcx) \Gamma h_v(0)|B(v) \rangle
&=~-\frac{\ii f_B M}{2}\;\text{Tr}\bigg\{ \gamma_5\,\Gamma \frac{1+v\sla}{2} 
\bigg( \Phi_+ - \frac{\lcx\sla}{2(v\lcx)}
\Big[\Phi_+-\Phi_-\Big]\bigg)\bigg\},
\\%
\label{Kodaira_vd}
(v\pd)\,\langle 0| \qb(x) \Gamma h_v(0)|B(v) \rangle\big|_{x=\lcx}
& = \frac{\ii f_B M}{2}\;\text{Tr}\bigg\{ 
 \gamma_5\,\Gamma \frac{1+v\sla}{2} 
\bigg[\frac{\lcx\sla}{(v\lcx)}\Delta_1 - \Delta_2
\bigg]\bigg\},
\\%
\label{Kodaira_dmu}
\pd_\mu\,\langle 0| \qb(x)\gamma^\mu \Gamma h_v(0)|B(v) \rangle\big|_{x=\lcx}
& = \frac{\ii f_B M}{2}\;\text{Tr}
 \bigg\{ 
 \gamma_5\,\Gamma \frac{1+v\sla}{2} 
\bigg[\frac{\lcx\sla}{(v\lcx)} \Delta_3 + \Delta_4 \bigg]\bigg\}, 
\end{align}
with%
\begin{align}
\Delta_1 &= \frac{1}{2}\bigg( (v\pd) - \frac{1}{(v\lcx)}\bigg)\big[\Phi_+ - \Phi_-\big]\,,
\\
\Delta_2 &= \frac{1}{2}\bigg((v\pd) - \frac{1}{(v\lcx)}\bigg)\big[\Phi_+ - \Phi_-\big]
+  \frac{1}{2}(v\pd)\big[\Phi_+ + \Phi_-\big] \,,
\\
\Delta_3 & =\frac{1}{2} \bigg( (v\pd) - \frac{1}{(v\lcx)}\bigg)\big[\Phi_+ - \Phi_-\big]
+ (v\lcx)\frac{\pd }{\pd x^2}\big[\Phi_+ + \Phi_-\big]\,,
\\
\Delta_4 &= \frac{\pd \Phi_-}{\pd(v\lcx)} - \frac{1}{(v\lcx)}\big[\Phi_+ - \Phi_-\big]\,, 
\end{align}
 where $\pd_\mu = v_\mu\,{\pd}/{\pd(v x)} + 2 x_\mu \,{\pd}/{\pd x^2}$ and, therefore, we have
$(v\pd)\, \Phi = {\pd\, \Phi}/{\pd(v x)} + 2 (vx) \,{\pd \,\Phi}/{\pd x^2}$.

Computing these traces it is easily seen that for the determination of $\Delta_1 \ldots \Delta_4$
only the Dirac structure $\Gamma= \gamma_5 \gamma_\alpha$ must be considered
(arguments $v\lcx$ omitted):
\begin{align}
 (v\pd)\,\langle 0| \qb(x) \gamma_5 \gamma_\alpha h_v(0)|B(v) \rangle\big|_{x=\lcx}
 =& ~\ii f_B M\,
 \left[ \lcx_\alpha\, \Delta_1 - v_\alpha (v\lcx) \Delta_2 \right]/(v\lcx),
 \label{Kodaira_bi2}
 \\
 \pd_\mu\,\langle 0| \qb(x)\gamma^\mu \gamma_5 \gamma_\alpha h_v(0)|B(v) \rangle\big|_{x=\lcx}
 =&~ \ii f_B M\,
 \left[  \lcx_\alpha\, \Delta_3 + v_\alpha (v\lcx)\Delta_4  \right]/(v\lcx)\,,
 \label{Kodaira_bi5}
\end{align} 

Analogously, the vacuum-to-meson matrix element containing a trilocal quark-antiquark-gluon operator is parametrized in terms of four three-particle LCDAs $\hat\Psi_A(v\lcx;\vartheta),\,\hat\Psi_V(v\lcx;\vartheta),\,\hat X_A(v\lcx;\vartheta)$ 
and $\hat Y_A(v\lcx;\vartheta)$ with $\vartheta$ being restricted to $0\leq\vartheta\leq 1$ 
as follows \cite{Kodaira2001,L05}:
\begin{align}
\langle 0 | \qb(\lcx) F_{\mu\nu}(\vartheta\lcx) \lcx^\nu \Gamma h_v(0) | B(v)\rangle
 ~&=~
  \frac{ f_B M}{2} \;
\Tr \bigg\{\gamma_5 \Gamma \frac{1+v\sla}{2}
\bigg( \big(v_\mu \lcx\sla - (v\lcx)\gamma_\mu\big) \left[\hat\Psi_A- \hat\Psi_V\right](v\lcx;\vartheta) 
\nonumber\\
&\hspace{1cm}
 -\ii \sigma_{\mu\nu} \lcx^\nu \hat\Psi_V(v\lcx;\vartheta) 
 - \lcx_\mu \hat X_A(v\lcx;\vartheta) 
 + \frac{\lcx_\mu\lcx\sla}{(v\lcx)}\hat Y_A(v\lcx;\vartheta) \bigg) \bigg\}\,,
 \label{Kodaira_tri0}
 \\
 \langle 0 | \qb(\lcx) v^\mu F_{\mu\nu}(\vartheta\lcx) \lcx^\nu \Gamma h_v(0) | B(v)\rangle
 ~&=~
  \frac{f_B M}{2} \;
\Tr \bigg\{\gamma_5 \Gamma \frac{1+v\sla}{2}
\big[ \lcx\sla \;\Theta_1 (v\lcx;\vartheta) - (v\lcx)\, \Theta_2 (v\lcx;\vartheta) \big]\bigg\},
\label{Kodaira_tri_v}
\\
 \langle 0 | \qb(\lcx) \gamma^\mu F_{\mu\nu}(\vartheta\lcx) \lcx^\nu \Gamma h_v(0) | B(v)\rangle
 ~&=~
  \frac{f_B M}{2} \;
\Tr \bigg\{\gamma_5 \Gamma \frac{1+v\sla}{2}
\big[ \lcx\sla \;\Theta_3(v\lcx;\vartheta)  + (v\lcx)\,\Theta_4(v\lcx;\vartheta) \big]\bigg\}. 
\label{Kodaira_tri_gamma}
\end{align}
Again, computing these traces for the Dirac structure $\Gamma= \gamma_5 \gamma_\alpha$ 
(arguments $v\lcx$ and $\vartheta$ omitted) one obtains:
\begin{align}
 \langle 0|\qb(\lcx)&v^\mu F_{\mu\nu}(\vartheta \lcx) \lcx^\nu \gamma_5 \gamma_\alpha h_v(0)|B(v) \rangle
~=~
 f_B M \left[  \lcx_\alpha \Theta_1 - v_\alpha (v\lcx)\Theta_2\right]\,, 
\label{TRI2}\\
 \langle 0|\qb(\lcx)&\gamma^\mu F_{\mu\nu}(\vartheta \lcx) \lcx^\nu \gamma_5 \gamma_\alpha h_v(0)|B(v) \rangle
~=~
f_B M \left[\lcx_\alpha \Theta_3 + v_\alpha (v\lcx)\Theta_4\right]\,, 
\label{TRI5}
\end{align}
with
\begin{align}
\label{T1}
\Theta_1 &= \big[\hat \Psi_A + \hat Y_A\big](v\lcx;\vartheta)\,,
\\
\label{T2}
\Theta_2 &= \big[\hat \Psi_A + \hat X_A\big](v\lcx;\vartheta)\,,
\\
\label{T3}
\Theta_3 &= \big[\hat \Psi_A + 2 \hat \Psi_V + \hat X_A\big](v\lcx;\vartheta)\,,
\\
\label{T4}
\Theta_4 &=2\big[\hat \Psi_A - \hat \Psi_V \big] (v\lcx;\vartheta)\,.
\end{align}

Putting expressions (\ref{Kodaira_bi2}) and (\ref{Kodaira_bi5}) and (\ref{TRI2}) and (\ref{TRI5}) into the equations of motion, (\ref{bt1}) and (\ref{bt2}), one arrives at the four differential equations of Ref.~\cite{Kodaira2001} connecting the 2-particle and 3-particle DAs. However, this approach has the drawback that the $x$-dependence of the distribution amplitudes, especially their dependence on $x^2$, is unknown.
In Ref.~\cite{Kodaira2001} this is partially circumvented by compensating $\pd \Phi_+/\pd x^2$ through combining two of the equations, disregarding another one containing also $\pd \Phi_-/\pd x^2$ and finally solving only two equations which are independent of these derivatives. Below, in Sect.~\ref{EoMKodaira} we show that, despite omitting part of the information, their result concerning the Mellin moments $\Phi_{\pm|n}(x^2=0)$ was complete. There, we also discuss the attempts \cite{Kodaira2003,HWZ05,HQW2006} to solve the problem of transverse momentum dependence of the B-meson wave function, i.e., to take into account the $x^2$-dependence of $\Phi_{\pm|n}(x^2\neq 0)$.
\smallskip

\noindent
{\it (2) General proof of sufficiency of the axial vector structure}\\
It is our aim to circumvent these drawbacks by using the decomposition of the appropriately parametrized 2- and 3-particle vacuum-to-meson matrix elements into {\it off-cone DAs of well-defined geometric twist whose $x$-dependence is completely known}. Introducing corresponding (double) Mellin moments the necessary differentiations can be easily performed. When restricting to the light-cone they simply lead to ordinary algebraic equations for the corresponding LCDAs of (even) twists $\tau=2,4$ and $6$. 

In principle, it is possible to restrict these considerations on the axial vector structure.
At first, Eq.~(\ref{Kodaira_bi2}) is sufficient to determine $\Delta_1$ and $\Delta_2$. But, looking at Eq.~(\ref{Kodaira_bi5}) one observes that, in order to compute $\Delta_3$ and $\Delta_4$, it seems to be necessary to know the twist decomposition of the second stage tensor operator or, having in mind relation (\ref{Sigma}), of the pseudo scalar and skew tensor operator. However, as will be shown now, the corresponding matrix elements can be related to that of the axial vector operator. 
 
Let us now derive the corresponding operator relations in the case of bilocal operators.
Using the identities (\ref{Chisholm}) and (\ref{Sigma}) one gets:
\begin{align}
\qb(x)\gamma^\mu \gamma_5 \gamma_\alpha \gamma_\beta h_v(0)
&= 
g_{\alpha\beta}\,\qb(x)\gamma^\mu \gamma_5 h_v(0)
-
\qb(x)\gamma^\mu \gamma_5 \ii \sigma_{\alpha\beta} h_v(0)
\nonumber \\ &
=-\,\big[ 
   {g^{\mu}}_{\alpha} {g_{\beta}}^{\rho}
 - {g^{\mu}}_{\beta} {g_{\alpha}}^{\rho}
 + g_{\alpha\beta} g^{\mu\rho} \big] 
 \, \qb(x) \gamma_5 \gamma_\rho h_v(0)
  - \ii {\epsilon_{\alpha\beta}}^{\mu\rho} 
 \,\qb(x) \gamma_\rho h_v(0).
\label{X}
\end{align}
Truncating both parts with $\pd_\mu$ one obtains a relation for 
$\Gamma = \gamma_{5}\ii \sigma_{\alpha\beta}$: 
\begin{align}
\partial_\mu\,&\qb(x) \gamma^\mu \gamma_5 \ii \sigma_{\alpha\beta} h_v(0)
=
 \partial_\alpha \,\qb(x) \gamma_5 \gamma_\beta h_v(0)
 -
 \partial_\beta \,\qb(x) \gamma_5 \gamma_\alpha h_v(0)
  + \ii {\epsilon_{\alpha\beta}}^{\rho\sigma} 
 \,\partial_\rho \,\qb(x) \gamma_\sigma h_v(0).
\label{X1}
\end{align}
Multiplying both sides of this equation with $v^\beta$ leads to a relation for 
$\Gamma = \gamma_{5}\gamma_{\alpha}$:
\begin{align}
\partial_\mu\, &\qb(x)\gamma^\mu \gamma_5\gamma_\alpha h_v(0)
=\big[
v^\mu g_{\alpha}^{\phantom{\alpha}\beta}- v^\beta g^\mu_{\phantom{\mu}\alpha} - 
v_\alpha g^{\mu\beta}
\big] \, \partial_\mu \,\qb(x) \gamma_5 \gamma_\beta h_v(0)
+ \ii \epsilon_{\alpha}^{\phantom{\alpha}\mu\rho\sigma}v_\rho \,\partial_\mu \,
\qb(x) \gamma_\sigma h_v(0) 
\nonumber \\
&=  (v \partial)\,\qb(x) \gamma_5 \gamma_\alpha h_v(0)
  - \partial_\alpha \;\qb(x) \gamma_5 h_v(0)
  + v_\alpha\; \partial_\mu \, \qb(x) \gamma^\mu \gamma_5 h_v(0)
  + \ii \epsilon_{\alpha}^{\phantom{\alpha}\mu\rho\sigma}v_\rho \, \partial_\mu \,
    \qb(x) \gamma_\sigma h_v(0) \,.
\label{X2}
\end{align}
However, multiplying subsequently with $v^\alpha$ does not lead to another independent relation
for $\Gamma = \gamma_{5}$.
Concerning the operator expressions on the LHS of Eq.~(\ref{bt2}) we may state also the
following relation, 
\begin{align}
&\qb(x) \gamma_5 \ii \sigma_{\alpha\beta} h_v(0)
= 
 \qb(x) \gamma_5 (v_\alpha \gamma_\beta - v_\beta \gamma_\alpha) h_v(0)
  - \ii {\epsilon_{\alpha\beta}}^{\rho\sigma} 
 \,v_\rho\,\qb(x) \gamma_\sigma h_v(0),
\label{X11}
\end{align}
which is obtained by multiplying (\ref{X}) with $v_\mu$. 

When taking vacuum-to-meson matrix elements of relations (\ref{X}) -- (\ref{X11})
the vector parts accompanying the $\epsilon$-tensor do not contribute
since $B$-mesons are pseudo-scalar, 
$\langle 0 |\qb(x)\gamma_\sigma h_v(0)|B(v) \rangle \equiv 0$; therefore we get  
\begin{align}
\partial_\mu\, \langle 0|\qb(x)\gamma^\mu \gamma_5\gamma_\alpha h_v(0)|B(v)\rangle
& = (v \partial)\,\langle 0|\qb(x) \gamma_5 \gamma_\alpha h_v(0)|B(v)\rangle
\nonumber \\& ~~~ 
	-\left(v_\alpha \partial_\mu 
	+ v_\mu \partial_\alpha \right)\langle 0|\qb(x) \gamma_5 \gamma^\mu  h_v(0)|B(v)\rangle\,, 
\label{X20}
\\
\pd_\mu\,\langle 0|\qb(x)\gamma^\mu\gamma_5\ii\sigma_{\alpha\beta} h_v(0)|B(v)\rangle
&=
 \pd_\alpha \,\langle 0|\qb(x) \gamma_5 \gamma_\beta h_v(0)|B(v)\rangle
 -
 \pd_\beta \,\langle 0|\qb(x) \gamma_5 \gamma_\alpha h_v(0)|B(v)\rangle\,,
 \label{X21}
\\
(v\pd)\,\langle 0|\qb(x)\gamma_5\ii\sigma_{\alpha\beta} h_v(0)|B(v)\rangle
&=(v\pd)\,
 \langle 0|\qb(x) \gamma_5 (v_\alpha \gamma_\beta 
 - v_\beta \gamma_\alpha) h_v(0)|B(v)\rangle\,,
 \label{X22}
\end{align}
as well as
\begin{align}
(v\partial)\, \langle 0|\qb(x) \gamma_5 h_v(0)|B(v)\rangle
& = 
v_\alpha\, (v\partial) \,\langle 0|\qb(x) \gamma_5 \gamma^\alpha  h_v(0)|B(v)\rangle
 \label{X23}
\end{align}
for the scalar operator.
Obviously, all these relations are valid only under the constraints of HQET. 

From these relations it becomes obvious that all the expressions $\Delta_1, \ldots, \Delta_4$
may be derived by the help of the first derivative of the axial vector DA alone, either using Eqs.~(\ref{Kodaira_bi2}) and (\ref{Kodaira_bi5}) or, eventuelly, using the structure of the pseudo scalar and skew tensor matrix elements.
Therefore, only  the geometric twist decomposition of $\langle 0 |\qb(x)\gamma_5 \gamma_\mu h_v(0)|B(v) \rangle$  has to be considered in detail. 

Concerning the tri-local operators and their matrix elements the situation is more difficult.
At first, relations similar to (\ref{X}) -- (\ref{X11}) also hold for the tri-local operators 
$\qb(x) F_{\mu\nu}(\vartheta x) x^\nu \gamma^\mu \Gamma h_v(0)$ and
$\qb(x) F_{\mu\nu}(\vartheta x) x^\nu \Gamma h_v(0)$ on the RHS of Eqs.~(\ref{bt1}) and (\ref{bt2}), respectively. But now, taking matrix elements, the $\epsilon$-terms, when multiplied with the vector part, $\ii \epsilon_{\alpha\beta}^{\phantom{\alpha\beta}\rho\sigma}\,
\langle 0|\qb(x) F_{\rho\nu}(\vartheta x) x^\nu \gamma_\sigma h_v(0)|B(v)\rangle$,
will contribute.

Furthermore, looking at Eq.~(\ref{TRI5}) we notice that only the twist decomposition of the axial vector LCDA is necessary to determine $\Theta_3$ and $\Theta_4$. However, the determination of 
$\Theta_1$ and $\Theta_2$ is much more involved. This becomes obvious considering the axial vector operator $\qb(x) F_{\mu\nu}(\vartheta x) x^\nu \gamma_5 h_v(0)$ which by using the trace formula (\ref{Kodaira_tri0}) differs in their structure from (\ref{TRI2}) but is simply given by $f_B M \lcx_\alpha \left[\Theta_1 - \Theta_2\right]$. Therefore, $v_\mu$ {\it has to be considered as an external parameter} being irrelevant for the twist decomposition which, for dimensional reasons, has to be parametrized by $x_\mu$ and not by $v_\mu$! So, the computation of $\Theta_1$ and $\Theta_2$ requires the determination of the twist decomposition of the second stage tensor 3-particle LCDA which can be decomposed in an antisymmetric and a symmetric part. Whereas the antisymmetric part is easily parametrized by $v_{[\alpha} \lcx_{\beta]}$, the parametrization of the symmetric part is troublesome, because $g_{\alpha\beta}(vx), v_\alpha v_\beta (vx), 
v_\alpha x_\beta + x_\alpha v_\beta$ as well as $x_\alpha x_\beta/(vx)$ may occur.
This will be undertaken in the next section.



\setcounter{equation}{0}
\section{Two- and Three-Particle Distribution Amplitudes of 
Definite Geometric Twist and Relevant Derivatives on the Light-Cone}
\label{EoMdisc}


In this section we collect all ingredients of the geometric twist decompositions of relevant tensor operators which, due to Eqs.~(\ref{Kodaira_bi2}) and (\ref{Kodaira_bi5}) and (\ref{TRI2}) and (\ref{TRI5}), are necessary as input in the equations of motion.
This will be done in terms of Mellin moments.
By the way, there are three different types of matrix elements we are concerned with, namely \\
~ $\bullet$  bilocal matrix elements: we only need their off-cone twist decompositions up to terms proportional to $x^2$, since -- after carrying out a single derivative -- all the higher order terms vanish when projected onto the light-cone ($\lcx^2 =0$),\\
~ $\bullet$  trilocal matrix elements: we only need their on-cone twist decomposition which for the structures $F_{\mu\nu}(\vartheta \lcx) \lcx^\nu$ as well as $F_{\mu\nu}(\vartheta \lcx) \gamma^\mu \lcx^\nu$ is known partly from our previous work \cite{Bmeson}, and finally,\\
~ $\bullet$ total translation terms  
which require a special treatment. 
\smallskip 

\noindent
(1) ~~{\em Derivatives of the two-particle axial vector distribution amplitudes on the light-cone}: 
\\
According to the foregoing considerations, especially taking into account the relation (\ref{X20}),
it is only necessary to consider the {\em axial vector} DA and restrict the derivatives to the light-cone. 
However, there exist two independent parametrizations by $v_\mu$ and by $x_\mu/(vx)$ which will be considered separately.

Let us first present the the off-cone (geometric) twist decomposition of the axial vector distribution amplitude, resulting from Eqs.~(\ref{TWVg},~\ref{TWVu}), in terms of Mellin moments according to the definition (\ref{BL}) 
(with the convention $\binom{j} {k}\equiv 0$ if either 
$k > j$ or $k < 0$):
\begin{align}
\langle 0|&\qb(x)\gamma_5\gamma_\alpha  h_v(0)|B(P)\rangle /(\ii\,f_B M )
= 
\sum_\tau \sum_{n=0}^\infty 
{\cal P}^{(\tau)\alpha^\prime}_{\alpha|n} \frac{(-\ii P x)^n}{n!} 
 \left( 
 v_{\alpha^\prime} \varphi_{A1|n}^{(\tau)} 
 + \frac{x_{\alpha^\prime}}{vx} \varphi_{A2|n}^{(\tau)}
 \right) 
\nonumber \\ & =~
\sum_{n=0}^\infty \frac{(-\ii P x)^n}{n!} 
\sum_{j=0}^{\left[\frac{n}{2}\right]} \frac{1}{j!}	
  \left(\frac{-x^2}{4(vx)^2}\right)^{\!\!j}\!\frac{n!}{(n-2j)!}	
	\sum_{k=0}^{j+1}(-1)^k \,\frac{(n-j-k)!}{(n+2-k)!}\,
	\nonumber\\
&\qquad \qquad\qquad \qquad
	\times\left[v_\alpha \,(n+1-j-k)\binom{j}{k} -  
	\frac{x_\alpha}{2 (vx)}\, (n-2j) \binom{j+1}{k}	\right]\!
	(n+2-2k)\,\varphi_{A1|n}^{(2+2k)}
	\label{Y1}
	\\
 &~+ \frac{x_{\alpha}}{vx}
\sum_{n=1}^\infty \frac{(-\ii P x)^n}{n!} 
\sum_{j=0}^{\left[\frac{n-1}{2}\right]} \frac{1}{j!}\!
  \left(\frac{-x^2}{4(vx)^2}\right)^{\!\!j}
	\sum_{k=0}^{j}\frac{(-1)^k \,(n-1)!\,(n-1-j-k)!}{(n-1-2j)!\,(n-k)!}
	\binom{j}{k} (n-2k)\,\varphi_{A2|n}^{(4+2k)}	,
	\label{Y1hat}
	\\
\mathrm{with}  
	& \qquad 
	\phantom{\varphi_{A2|n}^{(2)} \equiv 0 \qquad \mathrm{and} \qquad}
\varphi_{A1|n}^{(2+2j)} = 0 
\qquad \mathrm{for} \quad n < j\,,
\label{Y1a}
\\
  & \qquad
\varphi_{A2|n}^{(2)} \equiv 0 
\qquad \mathrm{and} \qquad
\varphi_{A2|n}^{(4+2j)} = 0 
\qquad \mathrm{ for} \quad n \leq j\,,
\label{Y1b}
\end{align}
resulting from the restrictions of the $j$-summation and being compatible with the $k$-summation up to $j+1$. Furthermore, it can be shown that, besides the terms $v_\alpha \varphi_{A1|n}^{(2)}$ which appear for $j=0$ at arbitrary $n$, the $v$-parametrized part of the axial vector DA depends only on the combinations $\varphi_{A1|n}^{(2+2k)} - \varphi_{A1|n}^{(4+2k)}$, 
whereas, besides of the terms $\varphi_{A2|n}^{(4)}$ for $j=0$, its $x$-parametrized part depends only on the combinations $\varphi_{A2|n}^{(4+2k)} - \varphi_{A2|n}^{(6+2k)}$.
Furthermore, in both cases, there occur no contributions of odd twist $\tau = 3 + 2j$.

The first derivative of the axial vector DA, when restricted to the light-cone, reads:
\begin{align}
\pd_\beta\,\langle 0|&\qb(x)\gamma_5\gamma_\alpha h_v(0)|B(P)\rangle\Big|_{x=\lcx}
= \ii f_B M \bigg\{
\nonumber \\ 
&\qquad
\sum_{n=1}^\infty \frac{(-\ii P \lcx)^n}{n!} \bigg(
\frac{v_\alpha v_\beta}{(v\lcx)} n \,\varphi_{A1|n}^{(2)}
-\frac{\delta_{\alpha\beta}(v\lcx)
 +(n-1)(v_\alpha \lcx_\beta + v_\beta \lcx_\alpha)}{(v\lcx)^2}
 \frac{n}{2(n+1)}
 \Big[\varphi_{A1|n}^{(2)}-\varphi_{A1|n}^{(4)}\Big] \bigg)
 \nonumber \\
 &\quad 
 +\sum_{n=2}^\infty \frac{(-\ii P \lcx)^n}{n!} 
\frac{\lcx_\alpha\lcx_\beta}{(v\lcx)^3} \frac{n-2}{4(n+1)}
\Big[(n-1)\,\varphi_{A1|n}^{(2)}
 -{2n}\,\varphi_{A1|n}^{(4)}+(n+1)\,\varphi_{A1|n}^{(6)} 
\Big]
\nonumber
\\
&\quad
+\sum_{n=1}^\infty \frac{(-\ii P \lcx)^n}{n!} 
\frac{\delta_{\alpha\beta}(v\lcx) + (n-1) \lcx_\alpha v_\beta}{(v\lcx)^2}\,
\varphi_{A2|n}^{(4)}		
- \frac{\lcx_{\alpha}\lcx_\beta}{(v\lcx)^3}
\sum_{n=2}^\infty \frac{(-\ii P \lcx)^n}{n!} \frac{n-2}{2}
\left[\varphi_{A2|n}^{(4)}	- \varphi_{A2|n}^{(6)}	\right]\!\bigg\}.
\label{bt5}
\end{align}
Surprisingly, the $v$-parametrized part, being related to LCDA $\varphi_{A1|n}^{(\tau)}$, is symmetric in $(\alpha \beta)$, whereas the $x$-parametrized part, being related to LCDA $\varphi_{A2|n}^{(\tau)}$, contains also a non-symmetric term.

Now, multiplying by $v^\beta$, from Eq.~(\ref{bt5}) we obtain 
\begin{align}
(v\partial)\, 
\langle 0|\qb(x)\gamma_5\gamma_\alpha h_v(0)&|B(P)\rangle\Big|_{x=\lcx}
= \ii f_B M \bigg\{
\frac{v_\alpha}{2(v\lcx)}
\sum_{n=1}^\infty \frac{(-\ii P \lcx)^n}{n!}\,
 \left(\frac{n}{n+1}
 \left[ (n+2)\,\varphi_{A|n}^{(2)}+ n\, \varphi_{A|n}^{(4)} \right]
 +2\,\varphi_{A2|n}^{(4)} \right)
\nonumber \\ 
&-\frac{\lcx_\alpha}{4(v\lcx)^2}
\sum_{n=2}^\infty \frac{(-\ii P \lcx)^n}{n!} 
 \bigg(\frac{n-1}{n+1}
\left[(n+2)\,\varphi_{A|n}^{(2)}+n\,\varphi_{A|n}^{(4)}\right] -
\left[n\,\varphi_{A|n}^{(4)}+(n-2)\,\varphi_{A|n}^{(6)}\right]
\nonumber \\ 
&\qquad\qquad\qquad\qquad\qquad\quad
-2 \left[ n\,\varphi_{A2|n}^{(4)} + (n-2)\,\varphi_{A2|n}^{(6)}\right]
\bigg)\bigg\},
\label{bt6}
\end{align}
and, multiplying by $-g^{\alpha\beta}$, we obtain, analogous to (\ref{Kodaira_bi5}),
\begin{align}
\frac{1}{\ii f_B M}\;\pd_\mu \langle 0|\qb(x)\gamma^\mu \gamma_5 h_v(0)|B(P)\rangle\Big|_{x=\lcx}
&=  \Delta_3  +  \Delta_4  
= -\,  \frac{1}{(v\lcx)}\sum_{n=1}^\infty \frac{(-\ii P \lcx)^n}{n!} 
\left\{ n \, \varphi_{A1|n}^{(4)}  + (n+3)\,\varphi_{A2|n}^{(4)} \right\} , 
\label{bt60}
\end{align}
whereas, using relation (\ref{X21}), we find (in terms of $\varphi_{A2|n}^{(\tau)}$ only!)
\begin{align}
\frac{1}{\ii f_B M}\; \pd_\mu \langle 0|\qb(x)\gamma^\mu\gamma_5\ii\sigma_{\alpha\beta} h_v(0)|B(P)\rangle\Big|_{x=\lcx}
 &= 
 \frac{2\,v_{[\alpha}\lcx_{\beta]}}{(v\lcx)}\,\Delta_3
= \frac{2\,v_{[\alpha} \lcx_{\beta]}}{(v\lcx)^2}
\sum_{n=2}^\infty \frac{(-\ii P \lcx)^n}{n!} 
(n-1)\,\varphi_{A2|n}^{(4)},
\label{X8}
\end{align}
where, here and in the following, we use the convention 
$a_{[\alpha} b_{\beta]} = \frac{1}{2}(a_{\alpha} b_{\beta} - a_{\beta} b_{\alpha} )$. 

Let us now express $\Delta_k, k = 1 ,\ldots ,4$ through the moments $\varphi_{Ai|n}^{(\tau)}, i = 1,2$  of the axial vector DA as they follow from Eqs.~(\ref{Kodaira_bi2}) and (\ref{bt6}) -- (\ref{X8}):
\begin{align}
\Delta_1 
&= - \frac{1}{4(v\lcx)}
\sum_{n=2}^\infty \frac{(-\ii P \lcx)^n}{n!} \bigg\{
 \frac{n-1}{n+1} 
 \left[(n+2)\,\varphi_{A1|n}^{(2)} + n\,\varphi_{A1|n}^{(4)}\right] 
 \nonumber\\
&~\qquad\qquad\qquad
-\left[n\,\varphi_{A1|n}^{(4)} + (n-2)\,\varphi_{A1|n}^{(6)}\right]
-2\left[ n\,\varphi_{A2|n}^{(4)} + (n-2)\,\varphi_{A2|n}^{(6)}\right]
\bigg\}\,,
\label{D1}
\\
\Delta_2 
&= -\frac{1}{2(v\lcx)}
\sum_{n=1}^\infty \frac{(-\ii P \lcx)^n}{n!} \left\{\frac{n}{n+1}
 \left[ (n+2)\,\varphi_{A1|n}^{(2)}+ n\, \varphi_{A1|n}^{(4)} \right]
 +2\,\varphi_{A2|n}^{(4)} \right\}\,,
\label{D2}
\\
\Delta_3 
 & = \frac{1}{(v\lcx)}
\sum_{n=2}^\infty \frac{(-\ii P \lcx)^n}{n!} (n-1)\,\varphi_{A2|n}^{(4)} \,,
\label{D3}
\\
\Delta_4 
&= -\,\frac{1}{(v\lcx)}
\sum_{n=1}^\infty \frac{(-\ii P \lcx)^n}{n!} \left[ n\,\varphi_{A1|n}^{(4)}
+ 2(n+1)\,\varphi_{A2|n}^{(4)} \right]\,.
\label{D4}
\end{align}

\noindent
(2) ~~ {\em The relevant three-particle light-cone distribution amplitudes} :\\
After having determined the relevant bilocal LCDAs we now proceed to look at the trilocal ones occuring in Eqs.~(\ref{TRI2}) and (\ref{TRI5}). Since in that case we are dealing with matrix elements on the light-cone, we can make use of the results already obtained in \cite{Bmeson} but, in addition, we also take into account those parametrizations which have not been considered there.
Unfortunately, the twist projection operators do not commute with the vector $v_\mu$, Therefore, before multiplying with $v_\mu$ we have to determine the twist decomposition of the matrix elements $\langle 0|\qb(\lcx)F_{\mu\nu}(\vartheta\lcx)\lcx^\nu\gamma_5\Gamma h_v(0)|B(v)\rangle$.
However, due to its structure the matrix element must vanish when multiplied with $x^\mu$.
Due to this, we are confronted with only one type of DAs in the cases of pseudo scalar and axial vector matrix elements and three types of DAs in the case of second stage tensor matrix element so that we get: 
\begin{align}
\langle 0|\qb(\lcx)&F_{\mu\nu}(\vartheta \lcx)\lcx^\nu\gamma_5 h_v(0)|B(v)\rangle/(f_B M)
 \label{Dtri_v}
 =
 \sum_\tau \sum_{n=0}^\infty 
 {\cal P}^{(\tau)\mu^\prime}_{\mu|n+1} \frac{(-\ii P x)^n}{n!} 
 \left( x_{\mu^\prime} - \frac{v_{\mu^\prime} x^2}{(vx)} \right) 
 \Upsilon_{P|n}^{(\tau)}(\vartheta)  \Big|_{x=\lcx}
\nonumber \\
& \qquad \qquad \qquad \qquad \qquad \qquad \qquad \qquad \;
=
\lcx_\mu
\sum_{n=0}^\infty \frac{(-\ii P\lcx)^{n}}{n!} \;
\Upsilon_{P|{n}}^{(5)}(\vartheta),
 \\
\langle 0|\qb(\lcx)&F_{\mu\nu}(\vartheta\lcx)\lcx^\nu\gamma_\alpha h_v(0)|B(v)\rangle/(f_B M)
= \sum_\tau
 \sum_{n=0}^\infty 
 {\cal P}^{(\tau)[\mu^\prime\alpha^\prime]}_{[\mu\alpha]|n+1}
 \frac{(-\ii P x)^{n}}{n!}\,\ii\, \epsilon_{\mu^\prime\alpha^\prime\kappa\lambda}\,
 v^\kappa x^\lambda \,\Upsilon_{V|{n}}^{(\tau)}(\vartheta)\Big|_{x=\lcx}
\nonumber
\\
&\qquad \qquad \qquad \qquad \qquad \qquad \qquad \qquad \; 
= \ii\, \epsilon_{\mu\alpha\kappa\lambda}\,v^\kappa \lcx^\lambda
\sum_{n=0}^\infty \frac{(-\ii P\lcx)^{n}}{n!} 
\Upsilon_{V|{n}}^{(4)}(\vartheta)\,,
\label{Dtri_t2d}
\\
\langle 0|\qb(\lcx)&F_{\mu\nu}(\vartheta \lcx) \lcx^\nu \gamma_5 \gamma_\alpha
h_v(0)|B(v) \rangle/(f_B M)
=
 \sum_\tau \sum_{n=0}^\infty \bigg\{
 {\cal P}^{(\tau)[\mu^\prime\alpha^\prime]}_{[\mu\alpha]|n+1} \frac{(-\ii P x)^n}{n!} 
  \,v_{[\mu^\prime} x_{\alpha^\prime]} \,\Upsilon_{T1|n}^{(\tau)}(\vartheta)
  \nonumber\\
 &\qquad 
 +{\cal P}^{(\tau)(\mu^\prime\alpha^\prime)}_{(\mu\alpha)|n+1} \frac{(-\ii P x)^n}{n!} 
 \bigg( 
 	v_{(\mu^\prime} x_{\alpha^\prime)}\,\Upsilon_{T2|n}^{(\tau)}(\vartheta)
  +\frac{x_{\mu^\prime} x_{\alpha^\prime}}{(vx)}\,\Upsilon_{T3|n}^{(\tau)}(\vartheta)
  + \frac{v_{\mu^\prime} v_{\alpha^\prime }x^2}{2(vx)} 
  \left[\Upsilon_{T1|n}^{(\tau)} - \Upsilon_{T2|n}^{(\tau)}\right](\vartheta)
 \nonumber\\
 &\qquad \qquad \qquad \qquad \qquad \quad
  - \f12\,(vx)\,g_{\mu^\prime \alpha^\prime} 
  \left[\Upsilon_{T1|n}^{(\tau)}(\vartheta) + \Upsilon_{T2|n}^{(\tau)}(\vartheta)
  + 2 \frac{x^2}{(vx)}\Upsilon_{T3|n}^{(\tau)}(\vartheta)
  \right]
  \bigg)\!
  \bigg\}\bigg|_{x=\lcx}
\nonumber \\
&\qquad
	=	\frac{1}{2}\sum_{n=0}^\infty \frac{(-\ii P\lcx)^{n}}{n!}
	\bigg\{ \!\! 
	\left(v_{\mu} \lcx_{\alpha} - (v\lcx)\,g_{\mu \alpha}\right)\! 		
  \left[\Upsilon_{T1|n}^{(4)} + \Upsilon_{T2|n}^{(4)}\right]\!(\vartheta)
  - \lcx_{\mu} v_{\alpha}
  \left[\Upsilon_{T1|n}^{(4)} - \Upsilon_{T2|n}^{(4)}\right]\!(\vartheta) \bigg\}
\nonumber\\
& \qquad \quad 
	+ \frac{1}{2}\sum_{n=1}^\infty \frac{(-\ii P\lcx)^{n}}{n!}
	\frac{\lcx_{\mu} \lcx_{\alpha}}{(v\lcx)} 
	\left(2\,\Upsilon_{T3|{n}}^{(6)}(\vartheta)
  -\frac{n}{ n+1}
	\left[\Upsilon_{T2|{n}}^{(4)} - \Upsilon_{T2|{n}}^{(6)}\right]\!(\vartheta)\right)\,. 
\label{F1}
\end{align}

We should remark that, contrary to the convention used in \cite{Bmeson}, here, in order to be able to compare with the results in \cite{Kodaira2001}, we are forced to define the tri-local matrix elements without the overall factor $\ii$, cf.~Eqs.~(I.3.8) -- (I.3.17). 
(As a convention formulas from Ref.~\cite{Bmeson} are indicated by writing "I" in front of the number being cited.) 
Of course, this does not change any of the results obtained in \cite{Bmeson} since there we only considered relations between two- and between three-particle LCDAs separately.

The twist decomposition of the pseudo scalar and axial vector case have been given to simplify later on some notations.
In the axial vector case, instead of $\Upsilon_{A1|n}^{(\tau)}$ and $\Upsilon_{A2|n}^{(\tau)}$, which have been introduced in \cite{Bmeson} as related to 
$v_{\mu} \lcx_{\alpha}$ and $\lcx_{\mu} v_{\alpha}$, respectively, we now introduced  
$\Upsilon_{T1|n}^{(\tau)} =  \Upsilon_{A1|n}^{(\tau)} + \Upsilon_{A2|n}^{(\tau)}$ and $\Upsilon_{T2|n}^{(\tau)} =  \Upsilon_{A1|n}^{(\tau)} - \Upsilon_{A2|n}^{(\tau)}$ related to antisymmetric and symmetric kinematic coefficients $v_{[\mu} \lcx_{\alpha]}$ and $v_{(\mu} \lcx_{\alpha)}$, respectively, as well as $\Upsilon_{T3|n}^{(\tau)}$ related to $\lcx_{\mu} \lcx_{\alpha}/(vx)$ because it seems to be more appropriate to define the DAs in accordance with the various structures of well-defined geometric twist.

Let us remark that all terms in the first equalities of (\ref{Dtri_v}) and (\ref{F1}) which contain $x^2$ can be ignored because, also after twist decomposition, they have a common factor $x^2$  and, hence, vanish on the light-cone.

Concerning the structure of the second stage tensor (\ref{F1}) we should point to the following: 
An {\em antisymmetric} second stage tensor contains, in principle, 4 different twist structures \cite{Joerg}, but three of them vanish identically for the single possible parametrization; on the light-cone the remaining one $\Upsilon_{T1|{n}}^{(4)}$ is of twist-4. On the contrary, a {\em symmetric} second stage tensor contains, in principle, 9 different twist structures \cite{Joerg} (for each of the various parametrizations!) whereby three, namely two of odd twist and one of even twist, vanish identically in our case. Therefore, the various LCDAs $\Upsilon_{Ti|{n}}^{(\tau)}, i=1,2,3$ for the symmetric tensor case should occur with an additional subindex. However, due to the necessary vanishing after multiplication with $x^\mu$ and after restriction to the light-cone that distinction can be removed. As a result we obtain contributions of twist-4 and twist-6 which are the leading terms of two independent infinite towers whose additional higher twist terms appear with nontrivial powers of $x^2$ and hence vanish. 

In addition, we remark that in the above list of three-particle LCDAs the skew tensor structure is missing. The reason is that up to now we are unable to construct the twist projection operator onto a tensor of third stage having mixed symmetry, i.e. being neither totally symmetric nor totally antisymmetric (cf.~Ref.~\cite{Bmeson}). However, due to the result of the preceding Section, we do not need it for the aim of the present paper. It would only act as a consistency check. 

Now, let us multiply with $v_\mu$ in order to get the twist decomposition of the relevant
tri-local matrix elements:
\begin{align}
\langle 0|\qb(\lcx)v^\mu F_{\mu\nu}(\vartheta \lcx)\lcx^\nu\gamma_5 &h_v(0)|B(v)\rangle/f_B M
 \label{Dtri_vv}
  = (v\lcx)(\Theta_1 - \Theta_2)
  = (v\lcx) \sum_{n=0}^\infty  \frac{(-\ii P x)^n}{n!}
 \Upsilon_{P|{n}}^{(5)}(\vartheta)\,, 
 \\
\langle 0|\qb(\lcx)v^\mu F_{\mu\nu}(\vartheta \lcx) \lcx^\nu \gamma_5 &\gamma_\alpha
h_v(0)|B(v) \rangle/f_B M
=
\sum_{n=0}^\infty \frac{(-\ii P\lcx)^{n}}{n!}
\bigg(
\f12\,\lcx_\alpha \! \left[\Upsilon_{T1|n}^{(4)} + \Upsilon_{T2|n}^{(4)} \right]\!(\vartheta)
- \,v_\alpha (v\lcx)\,\Upsilon_{T1|n}^{(4)}(\vartheta)\bigg)
\nonumber\\
&\qquad\qquad\quad +\f12\,\lcx_\alpha
\sum_{n=1}^\infty \frac{(-\ii P\lcx)^{n}}{n!}
  \left(2\,\Upsilon_{T3|n}^{(6)} (\vartheta) -\frac{n}{n+1}
  \left[\Upsilon_{T2|n}^{(4)} - \Upsilon_{T2|n}^{(6)} \right]\!(\vartheta)\!\right),
\end{align}
and, using the results of Ref.~\cite{Bmeson},
\begin{align}
\label{Dtri_g_v}
 \langle 0|\qb(\lcx) F_{\mu\nu}(\vartheta \lcx) \gamma^\mu \lcx^\nu
\gamma_5 \gamma_\alpha &h_v(0)|B(v) \rangle/f_B M
~=~ v_\alpha (v\lcx)
\sum_{n=0}^\infty \frac{(-\ii  P\lcx)^n}{n!}\,\Omega_{A1|{n}}^{(3)}(\vartheta) 
\nonumber\\
 & \qquad\quad
 +\f12\,\lcx_\alpha \sum_{n=1}^\infty \frac{(-\ii  P\lcx)^n}{n!} 
\left(2\,\Omega_{A2|n}^{(5)}(\vartheta)
- \frac{n}{n+1}\left[\, \Omega_{A1|{n}}^{(3)}
- \Omega_{A1|{n}}^{(5)}\right]\!(\vartheta)\! \right).
\end{align}

Let us now express the $\Theta_k, k = 1,\ldots,4$ through the moments $\Upsilon_{T\,i}, i = 1,2,3$
and $\Omega_{A\,i}, i = 1,2$, of the axial vector DA as they follow from Eqs.~(\ref{TRI2}) and (\ref{TRI5})
$\big[$ observing $\Upsilon_{T3|0}^{(6)} = 0 = \Omega_{A2|0}^{(5)} \big]$:
\begin{align}
\Theta_1 &= \frac{1}{2}\sum_{n=0}^\infty \frac{(-\ii P\lcx)^{n}}{n!}\left\{
\left[\Upsilon_{T1|n}^{(4)} + \Upsilon_{T2|n}^{(4)} \right]\!(\vartheta)
+ \left(\!2 \Upsilon_{T3|n}^{(6)} (\vartheta)
  -\frac{n}{n+1} \left[\Upsilon_{T2|n}^{(4)} - \Upsilon_{T2|n}^{(6)} \right]\!(\vartheta)
  \!\right)\!\right\},
\label{Th1}
\\
\Theta_2 &= \sum_{n=0}^\infty \frac{(-\ii P\lcx)^{n}}{n!}\;
\Upsilon_{T1|n}^{(4)}(\vartheta)\,,
\label{Th2}
\\
\Theta_3 &= \frac{1}{2}\sum_{n=0}^\infty \frac{(-\ii P\lcx)^{n}}{n!}
\left(2\,\Omega_{A2|n}^{(5)}(\vartheta)
- \frac{n}{n+1}\left[\, \Omega_{A1|{n}}^{(3)}
- \Omega_{A1|{n}}^{(5)}\right]\!(\vartheta)\! \right),
\label{Th3}
\\
\Theta_4 &= \sum_{n=0}^\infty \frac{(-\ii P\lcx)^{n}}{n!}\;
\Omega_{A1|{n}}^{(3)}(\vartheta)\,.
\label{Th4}
\end{align}

Comparing Eqs.~(\ref{Th1}) -- (\ref{Th4}) with Eqs.~(\ref{T1}) -- (\ref{T4}) we find 
\begin{align}
\Psi_{A|n} &= \f12 \left(\Upsilon_{T1|n}^{(4)} + \Upsilon_{T2|n}^{(4)} \right)
&\quad &\mathrm{for}\qquad n\geq 0\,,
\label{K1}
\\
X_{A|n} &= \f12 \left(\Upsilon_{T1|n}^{(4)} - \Upsilon_{T2|n}^{(4)} \right)
&\quad &\mathrm{for}\qquad n\geq 0\,,
\label{K2}
\\
Y_{A|n} &= \Upsilon_{T3|n}^{(6)} 
				- \frac{n}{2(n+1)}\left(\Upsilon_{T2|n}^{(4)} - \Upsilon_{T2|n}^{(6)} \right)
				&\quad &\mathrm{for}\qquad n\geq 1\,,
\label{K3}				
\\
\Psi_{V|n} &= \f12 \left(\Upsilon_{T1|n}^{(4)} + \Upsilon_{T2|n}^{(4)} 
							- \Omega_{A1|n}^{(3)} \right) 	
&\quad &\mathrm{for}\qquad n\geq 0\,,
\label{K4}
\end{align}
where the first three relations could be read off also by comparing Eq.~(\ref{F1}) 
with Eq.~(I.A.21) of Ref.~\cite{Bmeson}; the last relation is obtained by observing
$\Theta_{4|n} = \Omega_{A1|{n}}^{(3)} = 2 \Psi_{A|n} - 2 \Psi_{V|n}$.
In \cite{Bmeson}
we also showed  $\Psi_{V|n} \equiv \Upsilon_{V|n}^{(4)}$ 
and $Y_{A|n} - X_{A|n} = \Upsilon_{P|n}^{(5)}$. From this, and comparing the various expressions for $\Theta_{3|n}+\Theta_{4|n}$ we find 
\begin{align}
\Upsilon_{T3|n}^{(6)} 
				- \frac{n}{2(n+1)}\left(\Upsilon_{T2|n}^{(4)} - \Upsilon_{T2|n}^{(6)} \right)
				&= \f12 \left(\Upsilon_{T1|n}^{(4)} - \Upsilon_{T2|n}^{(4)} \right) + \Upsilon_{P|n}^{(5)}
				&\quad &\mathrm{for}\qquad n\geq 1\,,
\label{Z01}\\
\Upsilon_{T1|n}^{(4)} + \Upsilon_{T2|n}^{(4)} - \Omega_{A1|n}^{(3)} &= 2 \Upsilon_{V|n}^{(4)}
&\quad &\mathrm{for}\qquad n\geq 0\,,
\label{Z02}\\
\Omega_{A1|n}^{(3)}- \frac{n}{2(n+1)}\left( \Omega_{A1|{n}}^{(3)}
- \Omega_{A1|{n}}^{(5)}\right) + \Omega_{A2|n}^{(5)}
&= 2 \Upsilon_{T1|n}^{(4)} + \Upsilon_{T2|n}^{(4)}  
&\quad& \mathrm{for} \qquad n\geq 1\,,
\label{Z03}
\end{align}
which allows, by using $\Upsilon_{V|n}^{(4)}$ and $\Upsilon_{P|n}^{(5)}$, to circumvent the complicated expression for $\Theta_{1|n}$ and $\Theta_{3|n}$, and also to avoid $\Omega_{A1|n}^{(3)}$.

By the foregoing considerations we observe that the four independent distribution amplitudes can be chosen as $\Upsilon_{T1|n}^{(4)}, \Upsilon_{T2|n}^{(4)}, \Upsilon_{P|n}^{(5)} $ and 
$\Upsilon_{V|n}^{(4)} $. This allows to write the (double) Mellin moments of $\Theta_k$ as follows:
\begin{align}
\Theta_{1|n}(\vartheta) &= \Upsilon_{T1|n}^{(4)}(\vartheta) + \Upsilon_{P|n}^{(5)}(\vartheta)\,,
\label{Th1n}
\\
\Theta_{2|n}(\vartheta) &= 
\Upsilon_{T1|n}^{(4)}(\vartheta)\,,
\label{Th2n}
\\
\Theta_{3|n}(\vartheta) &= 
\Upsilon_{T1|n}^{(4)}(\vartheta) +  2 \Upsilon_{V|n}^{(4)}(\vartheta)\,,
\label{Th3n}
\\
\Theta_{4|n}(\vartheta) &= 
\Upsilon_{T1|n}^{(4)}(\vartheta) + \Upsilon_{T2|n}^{(4)}(\vartheta) - 2 \Upsilon_{V|n}^{(4)}(\vartheta)\,.
\label{Th4n}
\end{align}
Let us mention that, according to the definition (\ref{DoubleMoments}) of the $\vartheta$-moments, by comparing 
powers in $\vartheta$, the relations (\ref{K1}) -- (\ref{Th4n}) hold equally well for the double moments 
$\Upsilon_{\bullet|n,m}^{(\tau)}$ itself.

Finally, let us put together all those lower three-particle moments which vanish due to the foregoing considerations.
According to the definitions (\ref{Dtri_v}) -- (\ref{F1}) as well as relations (\ref{Z01}) -- (\ref{Z03})
we find
\begin{align}
\Upsilon_{Ti|0}^{(6)} &=0&   &\mathrm{for} \quad i = 2,3 \,,
\label{Y01}\\
\Omega_{Ai|0}^{(5)}  &= 0& &\mathrm{for} \quad i = 1,2 \,,
\label{Y02}\\
2\Upsilon_{T1|0}^{(4)} + \Upsilon_{T2|0}^{(4)} &= \Omega_{A1|0}^{(3)} ,&&
\label{Y03}\\
\Upsilon_{T1|0}^{(4)} - \Upsilon_{T2|0}^{(4)}&= -2\Upsilon_{P|0}^{(5)}\,.&& 
\label{Y04}
\end{align}
\smallskip

\noindent
(3) ~~ {\em Remark on the total translation parts} :\\
Finally, there is the last term in (\ref{bt2}), related to the total translation, to be considered.  Yet, it is not known how to deal with these total derivatives by the method of twist decomposition. For that reason we introduce an ``effective mass'' $\bar{\Lambda}=M-m_b$ in style of the first of Refs.~\cite{HQETRev} and simply define
\begin{align}
\ii v^\mu \delta_\mu^T \;\{ \langle 0|\qb(x) \Gamma h_v(0)|B(P) \rangle \} \big|_{x=\lcx} 
= \bar{\Lambda}\;
\langle 0|\qb(\lcx)\Gamma h_v(0)|B(P)\rangle\,.
\label{lambda}
\end{align}
In the limit of infinite heavy quark masses $\bar \Lambda$ goes to zero. 
Furthermore, the contributions of 
the total derivative $v^\mu \delta_\mu^T \;\{ \langle 0|\qb(x) \Gamma h_v(0)|B(P) \rangle \} \big|_{x=\lcx}$ may be assumed to be small. Furthermore, this substitution must be done only for the axial vector case, 
$v^\mu \delta_\mu^T \{ \langle 0|\qb(x) \gamma_5 \gamma_\alpha h_v(0)|B(P)\rangle\}
\big|_{x=\lcx} \to -\ii \bar{\Lambda}\, \langle 0|\qb(\lcx)\gamma_5\gamma_\alpha h_v(0)|B(P)\rangle$, 
whose twist decomposition already has been determined in the preceding Section.

\setcounter{equation}{0}
\section{Relations connecting two- and three-particle Mellin moments}
\label{EoMrelations}

Now we are ready to discuss in detail the relations (\ref{bt1}) and (\ref{bt2}) for the 
(only relevant) axial vector case which connect the matrix elements (\ref{Kodaira_bi5}) with (\ref{TRI5}) and (\ref{Kodaira_bi2}) with (\ref{TRI2}) in combination with the corresponding total translation term, respectively.
Namely, from Eq.~(\ref{bt1}) with $\Gamma = \gamma_5 \gamma_\alpha$, by applying the EoM we obtain
\begin{align}
\pd_\mu\,\langle 0|\qb(x)\gamma^\mu\gamma_5\gamma_\alpha h_v(0)|B(v)\rangle\big|_{x = \lcx}
&=~ 
\ii \! \int_0^1\!\!\d\vartheta\;\vartheta\,
\langle 0|\qb(\lcx) F_{\mu\nu}(\vartheta \lcx) \lcx^\nu\gamma^\mu\gamma_5\gamma_\alpha h_v(0)|B(v)\rangle, 
\label{X4a}\\
\intertext{whereas from Eq.~(\ref{bt2}) after applying the EoM and using relation (\ref{lambda}) we obtain}
(v\partial)\,\langle 0|\qb(x)\gamma_5\gamma_\alpha h_v(0)|B(v)\rangle\big|_{x = \lcx}
&=~ 
\ii \! \int_0^1\!\!\d\vartheta\;(\vartheta-1)\,
\langle 0|\qb(\lcx) F_{\mu\nu}(\vartheta \lcx) \lcx^\nu v^\mu \gamma_5\gamma_\alpha h_v(0)|B(v)\rangle 
 \nonumber\\
 & \quad  
- \ii \bar{\Lambda}\;\langle 0|\qb(\lcx)\gamma_5\gamma_\alpha h_v(0)|B(P)\rangle\,.
\label{X4}
\end{align}
Below we derive also a relation, connecting both the left hand sites of Eqs. (\ref{X4a}) and 
(\ref{X4}) without using the total translation part. This leads to another relation connecting bi- and tri-local matrix elements.

Due to the appearance of the derivative, the two-particle LCDAs contain an overall factor $1/(v\lcx)$ whereas the three-particle LCDAs contain an overall factor $(v\lcx)$ and the total translation terms due to (\ref{lambda}) do not have such an additional factor. Therefore, in order to derive the wanted relations, we have to compare equal powers of $(v\lcx)$ or, equivalently, powers of $(-\ii P\lcx)$.
\medskip

\noindent
(1) To begin with, we consider relation (\ref{X4a}) by writing down the $v_\alpha (v\lcx)$- and $\lcx_\alpha$- terms of relations (\ref{Kodaira_bi5}) and (\ref{TRI5}) separately:
\begin{align}
\!\!v_\alpha:~{M^2}&
\sum_{n=1}^\infty \frac{(-\ii P \lcx)^{n-1}}{n!} \left[\, n\,\varphi_{A1|n}^{(4)}
+ 2\,(n+1)\,\varphi_{A2|n}^{(4)} \right]
=
\sum_{n=0}^\infty \frac{(-\ii P\lcx)^{n+1}}{n!}
\int_0^1\!\!\d\vartheta\;\vartheta\,
\Omega_{A1|{n}}^{(3)}(\vartheta)\,,
\nonumber\\
\!\!\lcx_\alpha:~{M^2}& \sum_{n=2}^\infty \frac{(-\ii P \lcx)^{n-1}}{n!}\, (n-1)\,\varphi_{A2|n}^{(4)}
=
\sum_{n=1}^\infty \frac{(-\ii P\lcx)^{n+1}}{n!}
\int_0^1\!\!\d\vartheta\;\vartheta 
\left(\!\frac{n}{2(n+1)}\left[ \Omega_{A1|{n}}^{(3)}
- \Omega_{A1|{n}}^{(5)}\right] - \Omega_{A2|n}^{(5)}\!\right)\!(\vartheta),
\nonumber
\end{align}
where the relative factor $1/(v\lcx)^2$ between left and right hand side has been rewritten as $-M^2/(-\ii P\lcx )^2$.
Comparing equal powers of $(-\ii P\lcx)$ and using relations (\ref{Z02}) and (\ref{Z03}), we find (remember $\varphi_{A\,i|0}^{(4)} = 0$)
\begin{align}
{M^2}
 \Big[n\,\varphi_{A1|n}^{(4)}+ 2(n+1)\,\varphi_{A2|n}^{(4)} \Big]
&=
n(n-1)\! \int_0^1\!\!\d\vartheta\;\vartheta
\left(\Upsilon_{T1|n-2}^{(4)} + \Upsilon_{T2|n-2}^{(4)} - 2 \Upsilon_{V|n-2}^{(4)}\right)(\vartheta) 
&\mathrm{for}~ n\geq 1\,,
\label{R1} 
\\ %
- {M^2} (n-1)\,\varphi_{A2|n}^{(4)}
&=
n (n-1)\!\int_0^1\!\!\d\vartheta\;\vartheta 
\left(\Upsilon_{T1|n-2}^{(4)} + 2 \Upsilon_{V|n-2}^{(4)}\right)(\vartheta)
& \mathrm{for}~ n\geq 2\,.
\label{R2} 
\end{align}
The last equation can be used to replace $\varphi_{A2|n}^{(4)}$ in the first one to get
\begin{align}
{M^2}\,\varphi_{A1|n}^{(4)}
&=
\!\int_0^1\!\!\d\vartheta\,\vartheta 
\left(
(3n+1)\Upsilon_{T1|n-2}^{(4)} + (n-1)\Upsilon_{T2|n-2}^{(4)} 
+
2 (n+3) \Upsilon_{V|n-2}^{(4)}\right)\!(\vartheta)
& \qquad\mathrm{for}~ n\geq 2\,.
\label{R3} 
\end{align}
\medskip

\noindent
(2) Next, we consider relation (\ref{X4}) by writing down the $v_\alpha (v\lcx)$- and $\lcx_\alpha$- terms of relations (\ref{Kodaira_bi2}) and (\ref{TRI2}) separately:
\begin{align}
v_\alpha:\quad
\frac{M^2}{2}
\sum_{n=1}^\infty &\frac{(-\ii P \lcx)^{n-1}}{n!} \left\{\frac{n}{n+1}
 \left[ (n+2)\,\varphi_{A1|n}^{(2)}+ n\, \varphi_{A1|n}^{(4)} \right]
 +2\,\varphi_{A2|n}^{(4)} \right\}
 \nonumber\\
& =
\sum_{n=0}^\infty \frac{(-\ii P\lcx)^{n+1}}{n!}\;
\int_0^1\!\!\d\vartheta\;(\vartheta -1)\,
\Upsilon_{T1|n}^{(4)}(\vartheta)
+ \bar{\Lambda}\,M
\sum_{n=0}^\infty \frac{(-\ii P\lcx)^{n}}{n!}\varphi_{A1|n}^{(2)}\,,
\label{}
\\ 
\lcx_\alpha:\quad
 \frac{M^2}{2}
\sum_{n=2}^\infty& \frac{(-\ii P \lcx)^{n-1}}{n!} \bigg\{
 \frac{n-1}{n+1} 
 \left[(n+2)\,\varphi_{A1|n}^{(2)} + n\,\varphi_{A1|n}^{(4)}\right] 
 \nonumber\\
&~\qquad\qquad
-\left[n\,\varphi_{A1|n}^{(4)} + (n-2)\,\varphi_{A1|n}^{(6)}\right]
-2\left[ n\,\varphi_{A2|n}^{(4)} + (n-2)\,\varphi_{A2|n}^{(6)}\right]
\bigg\}
\nonumber\\
&  =
\sum_{n=0}^\infty \frac{(-\ii P\lcx)^{n+1}}{n!}
\int_0^1\!\!\d\vartheta\;(\vartheta -1)
\left(\Upsilon_{T1|n}^{(4)} + \Upsilon_{T2|n}^{(4)} 
  -\frac{n}{n+1} \left[\Upsilon_{T2|n}^{(4)} - \Upsilon_{T2|n}^{(6)} \right] 
  + 2\,\Upsilon_{T3|n}^{(6)} \!\right)\!(\vartheta)
  \nonumber\\&\quad
+\bar{\Lambda}\,M
 \sum_{n=1}^\infty \frac{(-\ii P\lcx)^{n}}{n!}\left(
 \frac{n}{n+1}\Big[\varphi_{A1|n}^{(2)}-\varphi_{A1|n}^{(4)}\Big]- 2\,\varphi_{A2|n}^{(4)}
 \right).
\label{}
\end{align}
Again, comparing equal powers of $(-\ii P\lcx)$ we obtain 
\begin{align}
&\frac{M^2}{2(n+1)}
 \left[ (n+2)\,\varphi_{A1|n}^{(2)}+ n\, \varphi_{A1|n}^{(4)} \right]
 +\frac{M^2}{n}\,\varphi_{A2|n}^{(4)}
- \bar{\Lambda}\,M \varphi_{A1|n-1}^{(2)}
=
(n-1)\int_0^1\!\!\d\vartheta\;(\vartheta -1)\,
\Upsilon_{T1|n-2}^{(4)}(\vartheta)\,,
\label{R4} 
\\
&\frac{M^2 }{2(n+1)}
\left[
(n-1)(n+2)\,\varphi_{A1|n}^{(2)}-2\,n\,\varphi_{A1|n}^{(4)}-(n+1)(n-2)\,\varphi_{A1|n}^{(6)}
\right]
 -
 M^2  \left[ n\,\varphi_{A2|n}^{(4)} + (n-2)\,\varphi_{A2|n}^{(6)}\right]
\label{R5}\\
&\quad -
\bar{\Lambda} M \left((n-1)\!
\left[\,\varphi_{A1|n-1}^{(2)} - \varphi_{A1|n-1}^{(4)}\right] - 2\,n\,\varphi_{A2|n-1}^{(4)}\right)
=
2n(n-1)\!\int_0^1\!\!\d\vartheta\,(\vartheta-1)\!
\left(\!
 \Upsilon_{T1|{n-2}}^{(4)} + \Upsilon_{P|{n-2}}^{(5)}\right)\!(\vartheta),
\nonumber
\end{align}
where both relations hold for $n\geq 2$, but the first one may be used also for $n= 1$.
\medskip

\noindent
(3) In addition, we remark that from Eq.~(\ref{bt1}) and relation (\ref{X2}) together with the corresponding one for 
$\qb(x) F_{\mu\nu}(\vartheta x) x^\nu \gamma^\mu \gamma_5 \gamma_\alpha h_v(0) =
v^\beta\qb(x) F_{\mu\nu}(\vartheta x) x^\nu\gamma^\mu\gamma_5\gamma_\alpha\gamma_\beta h_v(0)$
when using the identity (\ref{Chisholm}),
we get 
\begin{align}
(v \partial)\,\langle 0| \qb(x)\gamma_5\gamma_\alpha & h_v(0) |B(v)\rangle\big|_{x = \lcx}
=~ 
v_\mu\,\partial_\alpha \;\langle 0|\qb(x)\gamma_5\gamma^\mu h_v(0)|B(v)\rangle\big|_{x = \lcx} 
\nonumber\\
&+  \ii \! \int_0^1\!\!\d\vartheta\; \vartheta \;
 \langle 0|\qb(\lcx) F_{\mu\nu}(\vartheta \lcx) \lcx^\nu 
 \Big(\big[
 v^\mu g_{\alpha}^{\phantom{\alpha}\sigma}- v^\sigma g^\mu_{\phantom{\mu}\alpha} 
 \big]\gamma_5 + \ii \epsilon_{\alpha}^{\phantom{\alpha}\mu\rho\sigma}v_\rho \Big)
 \gamma_\sigma h_v(0)|B(v)\rangle 
\nonumber\\
& - v_\alpha\; \Big(
\partial_\mu \, \langle 0|\qb(x) \gamma^\mu \gamma_5 h_v(0)|B(v)\rangle
- \ii \! \int_0^1\!\!\d\vartheta\; \vartheta\;
\langle 0|\qb(x) F_{\mu\nu}(\vartheta x) x^\nu \gamma^\mu \gamma_5 h_v(0)|B(v)\rangle  
\Big)\Big|_{x = \lcx}\,.
\nonumber 
\end{align}
But, due to Eq.~(\ref{bt1}) with $\Gamma = \gamma_5$ the terms in the last line of that equation cancel each other and, using the on-shell constraint (\ref{onshell}) as well as (\ref{X4a}), we obtain a relation connecting two- and three-particle DAs without the total translation part
\begin{align}
(v \partial)\,\langle 0| \qb(x)\gamma_5\gamma_\alpha  h_v(0) |B(v)\rangle\big|_{x = \lcx}
&=~ 
\pd_\mu\,\langle 0|\qb(x)\gamma^\mu\gamma_5\gamma_\alpha h_v(0)|B(v)\rangle\big|_{x = \lcx}
+\pd_\alpha \,\langle 0|\qb(x)\gamma_5 h_v(0)|B(v)\rangle\big|_{x = \lcx} 
\nonumber\\
&\quad 
-  \ii \! \int_0^1\!\!\d\vartheta\; \vartheta \;
 \langle 0|\qb(\lcx) F_{\mu\nu}(\vartheta \lcx)\lcx^\nu 
 \left(g^\mu_{\phantom{\mu}\alpha} \gamma_5
 - \ii \epsilon_{\alpha}^{\phantom{\alpha}\mu\rho\sigma}v_\rho \gamma_\sigma
  \right) h_v(0)|B(v)\rangle \,.
   \label{X3}
   \end{align}
The tri-local terms are given by Eqs.~(\ref{Dtri_v}) and (\ref{Dtri_t2d}) which introduce the DAs 
$\Upsilon_{P|{n}}^{(5)}(\vartheta)$ and $\Upsilon_{V|{n}}^{(4)}(\vartheta)$, respectively, the derivatives 
of the axial vector part (on the l.h.s.) and of the tensor part (first term on the r.h.s.) are given by (\ref{Kodaira_bi5}) and (\ref{Kodaira_bi2}), respectively, whereas  
the derivative of the pseudoscalar part (second term on the r.h.s.) can be determined with the help of relation (\ref{bt5}), eventually using the on-shell condition, as follows:
\begin{align}
\pd_\alpha \,\langle 0|\qb(x)\gamma_5 h_v(0)&|B(v)\rangle\big|_{x = \lcx} 
=
\ii f_B M 
\bigg\{
\frac{v_\alpha}{2(v\lcx)}
\sum_{n=1}^\infty \frac{(-\ii  P\lcx)^{n}}{n!}
\left(\frac{n}{n+1} \left[\,(n+2)\varphi_{A1|n}^{(2)} + n \,\varphi_{A1|n}^{(4)}\right] 
+ 2 n \,\varphi_{A2|n}^{(4)}\right)
\nonumber\\
&~~~
-\frac{\lcx_\alpha}{4(v\lcx)^2}
\sum_{n=2}^\infty \frac{(-\ii  P\lcx)^{n}}{n!}
\bigg(\frac{n-1}{n+1}
\left[(n+2)\,\varphi_{A1|n}^{(2)} + n \,\varphi_{A1|n}^{(4)}\right]
- \left[ n\, \varphi_{A1|n}^{(4)} +(n-2)\, \varphi_{A1|n}^{(6)}\right] 
\nonumber\\
& \qquad\qquad\qquad\qquad\qquad\qquad
+ 2 (n-2)\left[\,\varphi_{A2|n}^{(4)} - \varphi_{A2|n}^{(6)}\right]\bigg)
\bigg\}\,.
\label{Z1} 
\end{align}
Let us observe that, due to the symmetry in $(\alpha\beta)$ for the terms 
$\varphi_{A1|n}^{(\tau)}$ in (\ref{bt5}) the last result differs from (\ref{bt6}) only in the terms $\varphi_{A2|n}^{(\tau)}$.

Now, putting together all these expressions we obtain, after cancellation of various terms and already omitting corresponding sums, the following very simple expressions:
\begin{align}
\lcx_\alpha:\quad
0 &= \int_0^1\!\!\d\vartheta\; \vartheta \,
\left(
\Upsilon_{P|{n}}^{(5)} -  2 \Upsilon_{V|{n}}^{(4)}\right)\!(\vartheta)\,,
\label{Z2}
\\
v_\alpha:\quad
0 &= M^2 \left( n\, \varphi_{A1|n}^{(4)} + (n+3) \,\varphi_{A2|n}^{(4)} \right)
+ 2\,n (n-1) \int_0^1\!\!\d\vartheta\; \vartheta \;\Upsilon_{V|{n-2}}^{(4)}(\vartheta)\,.
\label{Z3}
\\
\intertext{Comparing the last equation with the sum of Eqs.~(\ref{R1}) and (\ref{R2}) we find another relation between the trilocal DAs, namely}
0 &= \int_0^1\!\!\d\vartheta\; \vartheta 
\left(2\Upsilon_{T1|{n}}^{(4)} + \Upsilon_{T2|{n}}^{(4)}+2\Upsilon_{V|{n}}^{(4)}\right)\!(\vartheta)\,.
\label{Z4}
\end{align}
Obviously, this expression can be used to replace relation (\ref{Z3}).
Here we should remark that, contrary to ``strong'' relations (\ref{K1}) -- (\ref{K4}) and (\ref{Z01}) -- (\ref{Z03}), 
the relations (\ref{Z2}) and (\ref{Z4}) hold for the integrated DAs only.

At this stage we should remark that no further relation can be derived. 
One might think that relation (\ref{X3}), together with (\ref{X4}), can be used to express the total derivation part as follows:
\begin{align}
v^\mu \langle 0|\delta_\mu^T \{\qb(x)\gamma_5\gamma_\alpha h_v(0)\}|B(v)\rangle\big|_{x=\lcx}
&=~ 
\pd_\mu\,\langle 0|\qb(x)\gamma^\mu\gamma_5\gamma_\alpha h_v(0)|B(v)\rangle\big|_{x = \lcx}
+\pd_\alpha \,\langle 0|\qb(x)\gamma_5 h_v(0)|B(v)\rangle\big|_{x = \lcx} 
\nonumber\\&\quad 
-  \ii \! \int_0^1\!\!\d\vartheta\; \vartheta \;
 \langle 0|\qb(\lcx) F_{\mu\nu}(\vartheta \lcx)\lcx^\nu 
 \left(g^\mu_{\phantom{\mu}\alpha} \gamma_5
 - \ii \epsilon_{\alpha}^{\phantom{\alpha}\mu\rho\sigma}v_\rho \gamma_\sigma
  \right) h_v(0)|B(v)\rangle
  \nonumber\\
&\quad -\ii\! \int_0^1\!\!\d\vartheta\,(\vartheta-1)\,
\langle 0|\qb(\lcx) F_{\mu\nu}(\vartheta \lcx) \lcx^\nu v^\mu \gamma_5\gamma_\alpha h_v(0)|B(v)\rangle\,.
   \label{X3a}
\end{align}
However, due to relations (\ref{Z2}) and (\ref{Z3}) this expression falls back to the original EoM (\ref{X4}).
And, on the other hand, when using Eq.~(\ref{X2}) together with (\ref{X4a}) according to
\begin{align}
v^\mu \langle 0|\delta_\mu^T \{\qb(x)\gamma_5\gamma_\alpha h_v(0)\}|B(v)\rangle\big|_{x=\lcx}
&=~ 
\pd_\alpha \,\langle 0|\qb(x)\gamma_5 h_v(0)|B(v)\rangle\big|_{x = \lcx} 
\nonumber\\
&\quad 
+ \ii \! \int_0^1\!\!\d\vartheta\; \vartheta \;
 \langle 0|\qb(\lcx) F_{\mu\nu}(\vartheta \lcx)\lcx^\nu \gamma^\mu \gamma_5
 \left(\gamma_\alpha - v_\alpha \right) h_v(0)|B(v)\rangle
  \nonumber\\
&\quad -\ii\! \int_0^1\!\!\d\vartheta\,(\vartheta-1)\,
\langle 0|\qb(\lcx) F_{\mu\nu}(\vartheta \lcx) \lcx^\nu v^\mu \gamma_5\gamma_\alpha h_v(0)|B(v)\rangle\,,
   \label{X3b}
\end{align}
having in mind relation (\ref{lambda}), we are finally led to Eqs.~(\ref{R4}) and (\ref{R5}).

The main results of this section are relations (\ref{R1}) -- (\ref{R3}), (\ref{R4}) and (\ref{R5}),  (\ref{Z2}) and (\ref{Z4}) connecting the moments of two- to three-particle DAs of low twist. 
In addition, equations (\ref{K1}) -- (\ref{K4}), (\ref{Z01}) -- (\ref{Z03}) and (\ref{Z2}) and (\ref{Z4}) connect
the moments of various three-particle DAs.
In the next section we convert these local results from the (double) Mellin moments into the corresponding ones of their nonlocal DAs.

Let us emphasize that all these equations exhibit exact relations between the various contributions of geometric twist. Thereby, the leading contributions of twist $\tau=2$ occur only for two-particle distribution amplitudes. Higher twist contributions $(\tau =4,6)$ occur for both two- and three-particle distribution amplitudes. Thus, contributions of lowest twist are related to the lowest Fock state. These conclusions, of course, depend on the validity of the assumptions made by introducing the ``effective mass'' $\bar\Lambda$.

\setcounter{equation}{0}
\section{Relations for two- and three-particle Distribution Amplitudes}
\label{EoMnonlocalrelations}

In this section we first reformulate the restrictions of two-particle DAs, especially the
restrictions concerning their lower even moments, in terms of the non-local distribution amplitudes which can be 
interpreted as Burkhardt-Cottingham-like sum rules. 
Second, we reformulate the relations (\ref{R1}) -- (\ref{R3}), (\ref{R4}) and (\ref{R5}), connecting two- to three-particle LCDAs for low twist in terms of the non-local light-cone distribution amplitudes.
And, finally, we reformulate the local relations (\ref{K1}) -- (\ref{K4}), (\ref{Z01}) -- (\ref{Z03}) and (\ref{Z2}) 
and (\ref{Z4}) in terms of non-local three-particle LCDAs.
\medskip

\noindent
(1) {\it  Burkhardt-Cottingham-like sum rules for distribution amplitudes of lower twists}\\
First of all, let us translate the restrictions which hold for the lowest Mellin moments of the two- and three-particle distribution amplitudes into their non-local form.

From restrictions (\ref{Y1a}) and (\ref{Y1b}) for the {\em two-particle DAs} we immediately conclude the following:
 \begin{align}
 \int_0^1 \d u\, u^n\, \varphi_{A1}^{(2+2j)}(u) &= 0& 
 &\mathrm{for} \quad  0 \leq  n < j\,,
 \label{U1}
 \\
 \varphi_{A2}^{(2)}(u) \equiv 0 \qquad \mathrm{and} \qquad
 \int_0^1 \d u\, u^n\, \varphi_{A2}^{(4+2j)}(u) &= 0& 
 &\mathrm{for} \quad  0 \leq  n < j\,.
 \label{U2}
 \end{align}
Analogously, for the {\em three-particle DAs}, according to Eqs.~(\ref{Y01}) -- (\ref{Y04}), we obtain
\begin{align}
 \int_{0}^1 \!\!\D \underline{u}\;  \Upsilon_{T2}^{(6)}(u_1, u_2)
 =\int_{0}^1 \!\!\D \underline{u}\;  \Upsilon_{T3}^{(6)}(u_1, u_2)
 &= 0\,,
 \label{U3}
 \\
  \intertext{}
 \int_{0}^1 \!\!\D \underline{u}\;  \Omega_{A1}^{(5)}(u_1, u_2)
 = \int_{0}^1 \!\!\D \underline{u}\;  \Omega_{A2}^{(5)}(u_1, u_2)
 &= 0\,,
 \label{U4}
 \\
 \int_{0}^1 \!\!\D \underline{u}
 \left(2 \Upsilon_{T1}^{(4)}+ \Upsilon_{T2}^{(4)}- \Omega_{A1}^{(3)} \right)(u_1, u_2) &= 0  \,,
\label{U5}
\\
 \int_{0}^1 \!\!\D \underline{u}
 \left(2\Upsilon_{P}^{(5)} + \Upsilon_{T1}^{(4)}- \Upsilon_{T2}^{(4)} \right)(u_1, u_2) &= 0 \,.
\label{U6}
 \end{align}
Remind, that any odd twist DAs  $\varphi_{Ak}^{(1+2j)}(u)$ and $\Upsilon_{Tk}^{(1+2j)}(u_1, u_2)$ as well as any even twist DAs $\Omega_{Ak}^{(2+2j)}(u_1, u_2)$ vanish.
Obviously, these relations are Burkhardt-Cottingham-like sum rules for the axial vector distribution amplitudes.
\medskip
 
\noindent
(2) {\it Relations connecting two- and three-particle dispersion amplitudes of low twist}\\
The non-local version of the relations 
(\ref{R1}) -- (\ref{R3}), (\ref{R4}) and (\ref{R5}), (\ref{Z2}) and (\ref{Z3}), connecting two- to three-particle LCDAs, cannot be reduced to the level of DAs itself but
must be given in terms of integrations over the DAs. The reason is that these relations connect Mellin and double Mellin transforms. This is not the case for the lowest order of relations
(\ref{R1}), (\ref{R4})  and (\ref{Z3}) connecting only Mellin moments as well as (\ref{Z01}), (\ref{R2}) and (\ref{Z2}) connecting only double Mellin moments.

Let us demonstrate the method of derivation on the generic case 
\begin{align}
\frac{1}{(n-r+1)}\;\psi_{n}
= 
 \int_0^1\!\!\d\vartheta\; f(\vartheta) \,\Omega_{n-s} (\vartheta )
\quad {\rm for} \quad n\geq s\,.
\nonumber 
\end{align}	
Multiplying both sides of that equation by $(-\ii P x)^{n-s}/{(n-s)!}$,
using the integral representation 
\begin{align}
 \frac{1}{n-r+1}
 &=~\int_0^1\!\d \lambda\; \lambda^{n-r} 
 \qquad\qquad {\rm for} \qquad n\geq r,
 \label{Mellin1} 
\end{align}
together with the definition of Mellin (double) moments and summing up, we obtain 
\begin{align}
\sum_{n=s}^\infty\frac{(-\ii P x)^{n-s}}{(n-s)!}&
 \int_0^1\! \d u \;u^n\int_0^1\!\! \d\lambda  \,\lambda^{n-r}\,\psi(u)
 = 
 \int_0^1\!\!\d\vartheta\, f(\vartheta) \,
 \sum_{n=s}^\infty\frac{(-\ii P x)^{n-s}}{(n-s)!}
 \int_0^1\!\D \underline{u}\;(u_1+\vartheta u_2)^{n-s}\Omega(u_1,u_2).
\nonumber 
\end{align}
	Obviously, the summation on both sides results in exponential functions. The LHS can be 
rewritten by changing $u \lambda = \lambda'$, exchanging $\lambda'$- and $u$-integration and renaming the variables $(\lambda',u) \rightarrow (u,w)$. 
Analogously, the RHS can be rewritten by changing $u_2 \vartheta = \vartheta'$, exchanging $\vartheta'$- and $u_2$-integration and renaming the variables $(\vartheta',u_2) \rightarrow (u_2,w)$. 
By this procedure we arrive at
\begin{align}
 \int_0^1\!\! \d u \; e^{-\ii u P x}\,u^{s-r}
\int_u^1\! \frac{\d w }{w}\, w^r\,\psi(w)
= 
 \int_0^1\!\D \underline{u}\;e^{-\ii (u_1+ u_2) P x}
 \int_{u_2}^1 \frac{\d w}{w}\, f\left(\frac{u_2}{w}\right)\,\Omega(u_1,w).
\label{A2n}
\end{align}	
Let us mention that we {\em need not restrict to the light-cone} since the content of that formula does not depend on whether we write $x$ or $\lcx$. 

In applying this to Eq.~(\ref{R1}) we divide it by $n(n-1)$, observe 
$(n+1)/[n(n-1)] = 2/(n-1) - 1/n$ and finally get with $s=2$, $r=2$ resp. $r=1$, and $f(\vartheta)=\vartheta$:
\begin{align}
\hspace{-.25cm}
M^2 \!\! \int_0^1\!\! \d u \; e^{-\ii u P x}\!
\int_u^1\!\! \d w \,w \!\left[\varphi_{A1}^{(4)}(w) + 2\left(2-\frac{u}{w}\right) \varphi_{A2}^{(4)}(w)\right]
= 
 \int_0^1\!\!\D \underline{u}\;e^{-\ii (u_1+ u_2) P x}\!
 \int_{u_2}^1 \!\frac{\d w}{w}\, \frac{u_2}{w}\,\Omega_{A1}^{(3)}(u_1,w),
\label{R1n}
\end{align}	%
and for relation (\ref{R2}), taking into account Eq.~(\ref{Z03}), we obtain
\begin{align}
\hspace{-.25cm}
M^2 \!\! \int_0^1\!\! \d u \,u \; e^{-\ii u P x}\!
&\int_u^1\!\! \d w  \, \varphi_{A2}^{(4)}(w)
= 
 \int_0^1\!\!\D \underline{u}\;e^{-\ii (u_1+ u_2) P x}\!
 \int_{u_2}^1 \!\frac{\d w}{w}\, \frac{u_2}{w} 
 \left[\,2\Upsilon_{T1}^{(4)}+\Upsilon_{T2}^{(4)}-\Omega_{A1}^{(3)}\right]\!(u_1,w),
\label{R2n}
\end{align}
where $\Omega_{A1}^{(3)}= \Upsilon_{T1}^{(4)}+ \Upsilon_{T2}^{(4)} - 2\Upsilon_{V}^{(4)}$ as given by Eq.~(\ref{Z02}).

Now, we consider relation (\ref{R4}). 
For $n=1$ it relates only two-particle DAs and is almost trivial,
\begin{align}
M \int_0^1\!\! \d u\;u\-\left(\left[3\,\varphi_{A1}^{(2)}(u) + \varphi_{A1}^{(4)}(u)\right] 
+ 4 \,\varphi_{A2}^{(4)}(u)\right) 
- 4\,\bar\Lambda \int_0^1\!\! \d u\;\varphi_{A1}^{(2)}(u) = 0\,,
\label{R4an}
\end{align}
whereas for $n\geq 2$ we find the following non-local relations,
\begin{align}
\int_0^1\!\! \d u \; e^{-\ii u P x}\!
\int_u^1\! \frac{\d w}{w} &
\bigg\{
\f12\, M^2 \!\left[(w^2-u^2)\,\varphi_{A1}^{(2)}(w)
+(w^2+u^2)\,\varphi_{A1}^{(4)}(w) + 2w(w-u)\varphi_{A2}^{(4)}(w)\right]\!
-\bar\Lambda M\, w \,\varphi_{A1}^{(2)}(w)
\bigg\}
\nonumber\\
&\qquad
= \int_0^1\!\D \underline{u}\;e^{-\ii (u_1+ u_2) P x}
 \int_{u_2}^1 \frac{\d w}{w}\, \left(\frac{u_2}{w}-1\right)
\Upsilon_{T1}^{(4)}(u_1,w) \,.
\label{R4bn}
\end{align}
The non-local version of relation (\ref{R5}) reads
\begin{align}
\int_0^1\!\! \d u \; e^{-\ii u P x}&
\int_u^1\! \frac{\d w}{w} 
\bigg\{\!
\f12\,M^2 \bigg(u(2w-u)\left[\varphi_{A1}^{(2)}(w)- \varphi_{A1}^{(4)}(w)\right]
-w(w-2u)\left[\varphi_{A1}^{(4)}(w)- \varphi_{A1}^{(6)}(w)\right]
\nonumber\\
&\qquad\quad 
-2w(w-2u)\left[\varphi_{A2}^{(4)}(w)- \varphi_{A2}^{(6)}(w)\right]\bigg)
-\bar\Lambda M\bigg(u \left[\varphi_{A1}^{(2)}(w)- \varphi_{A1}^{(4)}(w)\right]
- 2 w \varphi_{A2}^{(4)}(w)\bigg)\!
\bigg\}
\nonumber\\
&\qquad\quad 
= \int_0^1\!\D \underline{u}\;e^{-\ii (u_1+ u_2) P x}
 \int_{u_2}^1 \frac{\d w}{w} \left(\frac{u_2}{w}-1\right)
\left[\Upsilon_{T1}^{(4)}(u_1,w) +\Upsilon_{P}^{(5)}(u_1,w) \right].
\label{R5n}
\end{align}
\medskip

\noindent
(3) {\it Relations between three-particle distribution amplitudes of low twist}\\	
First, let us consider relations (\ref{Z2}) and (\ref{Z4}). Taking into account the steps leading to RHS of Eq.~(\ref{A2n}), they may be non-locally rewritten as
\begin{align}
0&= \int_0^1\D \underline{u}\;e^{-\ii (u_1+ u_2) P x}\int_{u_2}^1 \frac{\d w}{w} \frac{u_2}{w}
\left(\Upsilon_{P}^{(5)} - 2 \Upsilon_{V}^{(4)}\right)\!(u_1,w)\,,
\label{Z2nl}\\
0&= \int_0^1\D \underline{u}\;e^{-\ii (u_1+ u_2) P x}\int_{u_2}^1 \frac{\d w}{w} \frac{u_2}{w}
\left(2\Upsilon_{T1}^{(4)}+\Upsilon_{T2}^{(4)} + 2 \Upsilon_{V}^{(4)}\right)\!(u_1,w)\,.
\label{Z4nl}
\end{align}

Second, concerning relation (\ref{Z02}), the simplest of relations (\ref{Z01}) -- (\ref{Z03}), we get 
\begin{align}
\Omega_{A1}^{(3)}(u_1,u_2)
&= \left(\Upsilon_{T1}^{(4)} + \Upsilon_{T2}^{(4)} - 2\Upsilon_{V}^{(4)} \right)(u_1,u_2)\,.	
\label{KZ02nl}
\end{align}

Furthermore, the nonlocal version of (\ref{Z01}) may be obtained as follows:
\begin{align}
0 = &
 \sum_{n=0}^\infty\frac{(-\ii P x)^{n}}{n!}
 \int_0^1\!\D \underline{u}\;(u_1+\vartheta u_2)^{n}
\bigg\{
2\Upsilon_{P}^{(5)}(u_1,u_2)
\nonumber\\
&\qquad\qquad
+ \left[\Upsilon_{T1}^{(4)} - \Upsilon_{T2}^{(4)} 
- 2  \Upsilon_{T3}^{(6)} \right](u_1,u_2)
+ \Big( 1 - \int_0^1 \d\lambda\,\lambda^{n} \Big)
\left[\Upsilon_{T2}^{(4)} - \Upsilon_{T2}^{(6)} \right](u_1,u_2) \bigg\}
\nonumber\\
= & 
\int_0^1\!\d u_1\;\int_0^1\!\d u_2\;e^{-\ii (u_1+ \vartheta u_2) P x}
\left[2\Upsilon_{P}^{(5)} +\Upsilon_{T1}^{(4)} + \Upsilon_{T2}^{(6)} 
- 2  \Upsilon_{T3}^{(6)} \right](u_1,u_2)
\nonumber\\
& - \int_0^1 \d\lambda\,\int_0^1\!\d u_1\;\int_0^1\!\d u_2\;e^{-\ii (u_1+ \vartheta u_2)\lambda P x}
\left[\Upsilon_{T2}^{(4)} - \Upsilon_{T2}^{(6)} \right](u_1,u_2)\,,
\intertext{which, when integrated over $\vartheta$ with an arbitrary function $f(\vartheta)$ as}
&
\int_0^1\!\D \underline{u}\;e^{-\ii (u_1+ u_2) P x}
 \int_{u_2}^1 \frac{\d w}{w}\, f\left(\frac{u_2}{w}\right)\,
 \bigg\{\!
 \left[\,\Upsilon_{P}^{(5)} + \Upsilon_{T1}^{(4)}\right]
 - \f12\,\left[\Upsilon_{T1}^{(4)} - \Upsilon_{T2}^{(6)} 
 + 2 \Upsilon_{T3}^{(6)} \right]\bigg\}(u_1,w)
 \label{Omega5n}
\nonumber\\
& \qquad\qquad\qquad\qquad
=\f12 \int_0^1 \d\lambda\,
\int_0^1\!\D \underline{u}\;e^{-\ii (u_1+ u_2)\lambda P x}
 \int_{u_2}^1 \frac{\d w}{w}\, f\left(\frac{u_2}{w}\right)\,
\left[\Upsilon_{T2}^{(4)} - \Upsilon_{T2}^{(6)} \right](u_1,w)\,,
\end{align}%
may be used to replace $\left(\Upsilon_{P}^{(5)} + \Upsilon_{T1}^{(4)}\right)(u_1,w)$ in relation (\ref{R5n}).
Furthermore, introducing new variables $0 \leq u = u_1 + u_2 \leq 2$ and $\lambda' =  u \lambda$, then exchanging integrations over $u$ and $\lambda'$ and renaming thereafter these variables as $v$ and $u$, we may rewrite the r.h.s. 
of Eq.~(\ref{Omega5n}) as follows,
\begin{align}
\f12 \int_0^1\!\!\D \underline{u}\;e^{-\ii (u_1+ u_2) P x} 
 \int_{u_2}^1 \frac{\d w}{w}\, f\left(\frac{u_2}{w}\right)\,
 \int_{u_1}^{1}\frac{\d v}{v+u_2}
\left[\Upsilon_{T2}^{(4)} - \Upsilon_{T2}^{(6)} \right]\!(v,w)\,,
\label{Omega5n0}
\end{align}
where the support restriction $0\leq v \leq 1$ has been used for the $v$-integration. 
Now, since $f(\vartheta)$ is an arbitrary function and the exponentials build up a complete orthonormal system of functions we conclude that
\begin{align}
0 = \left[\, 2 \Upsilon_{P}^{(5)} + \Upsilon_{T1}^{(4)} + \Upsilon_{T2}^{(6)} 
- 2  \Upsilon_{T3}^{(6)} \right]\!(u_1,u_2)
-  \int_{u_1}^{1}\frac{\d v}{v+u_2}
& \left[\Upsilon_{T2}^{(4)} - \Upsilon_{T2}^{(6)} \right]\!(v,u_2)\,.
\label{Omega5n0l}
\end{align} 

Analogous to (\ref{Omega5n}) the $\vartheta$-integrated relation (\ref{Z03}) can be reformulated non-locally as follows:
\begin{align}
\int_0^1\!\D \underline{u}\;e^{-\ii (u_1+ u_2) P x}&
 \int_{u_2}^1 \frac{\d w}{w}\, f\left(\frac{u_2}{w}\right)\,
 \bigg\{\!
 \left[\,2 \Upsilon_{T1}^{(4)} + \Upsilon_{T2}^{(4)}  - \Omega_{A1}^{(3)}\right]
+ \left(\f12 \left[ \Omega_{A1}^{(3)} - \Omega_{A1}^{(5)}\right]
- \Omega_{A2}^{(5)}\right) \bigg\}(u_1,w)
\nonumber\\
& =\f12 \int_0^1 \d\lambda\,
\int_0^1\!\D \underline{u}\;e^{-\ii (u_1+ u_2)\lambda P x}
 \int_{u_2}^1 \frac{\d w}{w}\, f\left(\frac{u_2}{w}\right)\,
\left[\Omega_{A1}^{(3)} - \Omega_{A1}^{(5)}\right](u_1,w)\,,
\label{R2n0}
\end{align}%
leading to the relation
\begin{align}
0 = \left[\, 2 \Upsilon_{T1}^{(4)} + \Upsilon_{T2}^{(4)}  
- \f12 \left( \Omega_{A1}^{(3)} + \Omega_{A1}^{(5)}\right) - \Omega_{A2}^{(5)} 
\right]\!(u_1,u_2)
-  \f12 \int_{u_1}^{1}\frac{\d v}{v+u_2}
& \left[\Omega_{A1}^{(3)} - \Omega_{A1}^{(5)} \right]\!(v,u_2)\,.
\label{R2n0l}
\end{align} 

For the special case of zeroth double moments, i.e. for $n=0$, instead of 
(\ref{Omega5n0}) and (\ref{R2n0}) one gets
\begin{align}
\int_0^1\!\D \underline{u}
\left[\,\Upsilon_{P}^{(5)}
+ \f12\left(\Upsilon_{T1}^{(4)} - \Upsilon_{T2}^{(4)}\right) 
-   \Upsilon_{T3}^{(6)}\right]\!(u_1,u_2)= 0\,,
\label{Omega50}
\\
\int_0^1\!\D \underline{u}
\left[\, 2 \Upsilon_{T1}^{(4)} + \Upsilon_{T2}^{(4)}  
-  \Omega_{A1}^{(3)}  - \Omega_{A2}^{(5)} \right]\!(u_1,u_2)= 0\,,
\end{align}
which, of course, is consistent with relations (\ref{U3}) -- (\ref{U6}).


\setcounter{equation}{0}
\section{Connection with the work of Kawamura et al. and Huang et al.}
\label{EoMKodaira}

This section is devoted to relate the two- and three-particle distribution amplitudes of definite geometric twist and their Mellin (double) momenta given above to the corresponding ones of Ref.~\cite{Kodaira2001,Kodaira2003} and \cite{HWZ05,HQW2006}.
\medskip

\noindent
{\it (1) Connection of the DAs $\Phi_\pm(vx,x^2)$ with the DAs $\varphi_{A1|n}^{(\tau)}$ and ${\varphi}_{A2|n}^{(\tau)}$
of geometric twist $\tau$}\\ 
For our aim it is sufficient, to consider 
the axial vector part only and to do that up to terms of order $ {\cal O}(x^2)$. Taking 
into account the expressions (\ref{Y1}) and (\ref{Y1hat}) and definition (\ref{Kodaira_bi}),
\begin{align}
\langle 0| \qb(x) \gamma_5 \gamma_\alpha h_v(0)|B(v) \rangle
 &= ~
 \ii f_B M \!
 \left[\frac{x_\alpha}{2(vx)} \big[\, \Phi_+ - \Phi_- \big]
 - v_\alpha \Phi_+ \right],
 \nonumber
\end{align}
we obtain $\Phi_\pm(vx,x^2)$, as well as their difference, up to terms of order $x^2$ as follows:
\begin{align}
\left[\Phi_{+}-\Phi_{-}\right](vx,x^2)
 &=
- \sum_{n=0}^\infty \frac{(-\ii P x)^n}{n!} \left\{
 \frac{n}{n+1}\left[\varphi_{A1|n}^{(2)}- \varphi_{A1|n}^{(4)}\right]
 - 2\,{\varphi}_{A2|n}^{(4)}\right\}
\nonumber\\
&~~~
- \frac{x^2}{4(v x)^2}\sum_{n=2}^\infty \frac{(-\ii P x)^n}{n!} 
(n-2)\bigg\{
\frac{n-1}{n+1} \left[\varphi_{A1|n}^{(2)}- \varphi_{A1|n}^{(4)}\right]
\nonumber\\
&~~~\qquad\qquad\qquad\qquad\qquad
- 
\left[
{\varphi}_{A1|n}^{(4)}	- {\varphi}_{A1|n}^{(6)}\right]	
+ 2 \left[{\varphi}_{A2|n}^{(4)}	- {\varphi}_{A2|n}^{(6)}\right]\!
\bigg\}
\label{pml}\\
\intertext{}
&=
-\int_0^1 \d u \,e^{-\ii u Px} 
\bigg\{\!\left[\varphi_{A1}^{(2)} - \varphi_{A1}^{(4)}\right]\!(u)
- \int_u^1 \frac{\d w}{w}
\left[\varphi_{A1}^{(2)} - \varphi_{A1}^{(4)} \right]\!(w) - 2\varphi_{A2}^{(4)}(u)
\nonumber\\
&~~\qquad
- \frac{x^2M^2}{4}\int_u^1\!\! {\d w}\left(2w-3u\right)\frac{u}{w}
\left[\varphi_{A1}^{(2)} - \varphi_{A1}^{(4)} \right]\!(w)
\nonumber\\
&~~\qquad
-\frac{x^2M^2}{4} 
\int_u^1\!\! {\d w}\left(w-2u\right)
\bigg(\!\left[\varphi_{A1}^{(4)} - \varphi_{A1}^{(6)} \right]\!(w)
-2\left[\varphi_{A2}^{(4)} - \varphi_{A2}^{(6)}\right]\!(w)\bigg)\!
\bigg\},
\label{pmnl}\\
\Phi_{+}(vx,x^2)
 &=
 -\sum_{n=0}^\infty \frac{(-\ii P x)^n}{n!} 
 \left\{\varphi_{A1|n}^{(2)}
 -\frac{ x^2}{(v x)^2}\frac{n(n-1)}{4(n+1)} 
 \left[\varphi_{A1|n}^{(2)}- \varphi_{A1|n}^{(4)}\right]\right\}
 \label{pl}\\
 &=
-\int_0^1 \d u \,e^{-\ii u Px} 
\left\{
\varphi_{A1}^{(2)}(u) 
+ \frac{x^2M^2}{4}\; u^2 \int_u^1 \frac{\d w}{w}
\left[\varphi_{A1}^{(2)} - \varphi_{A1}^{(4)} \right](w)
\right\},
 \label{pnl}\\
 \Phi_{-}(vx,x^2)
 &=
 -\sum_{n=0}^\infty \frac{(-\ii P x)^n}{n!} \left\{
 \frac{1}{n+1}\left[\varphi_{A1|n}^{(2)}- \varphi_{A1|n}^{(4)}\right]
 +   \varphi_{A1|n}^{(4)}
 + 2\,{\varphi}_{A2|n}^{(4)}\right\}
\nonumber\\
&~~~
+ \frac{x^2}{4(v x)^2}
\sum_{n=2}^\infty \frac{(-\ii P x)^n}{n!} 
\bigg\{
\frac{2(n-1)^2}{(n+1)} \left[\varphi_{A1|n}^{(2)}- \varphi_{A1|n}^{(4)}\right]
\nonumber\\
&~~~\qquad\qquad\qquad\qquad\qquad
- 
(n-2)\bigg(
\left[{\varphi}_{A1|n}^{(4)}	- {\varphi}_{A1|n}^{(6)}\right]	
-
2 \left[{\varphi}_{A2|n}^{(4)}	- {\varphi}_{A2|n}^{(6)}	\right]\bigg)\!
\bigg\}
\label{ml}\\
&=
-\int_0^1 \d u \,e^{-\ii u Px} 
\bigg\{\int_u^1 \frac{\d w}{w}
\left[\varphi_{A1}^{(2)} - \varphi_{A1}^{(4)} \right](w)
+ \varphi_{A1}^{(4)}(u) + 2\, \varphi_{A2}^{(4)}(u)\bigg\}
\nonumber\\
&~~~ 
+\frac{x^2M^2}{4}
\int_0^1 \d u \,e^{-\ii u Px} 
\int_u^1 \frac{{\d w}}{w}\left(w-2u\right)
\bigg\{
2u\left[\varphi_{A1}^{(2)} - \varphi_{A1}^{(4)} \right]\!(w)
\nonumber\\
&~~\qquad\qquad\qquad\qquad\qquad
-w \bigg(\left[\varphi_{A1}^{(4)} - \varphi_{A1}^{(6)} \right]\!(w)
-\,2\left[\varphi_{A2}^{(4)} - \varphi_{A2}^{(6)}\right]\!(w)\bigg)
\bigg\}\,,
\label{mnl}
\end{align}
where the sums go over the Mellin moments of $\Phi_{\pm|n}(x^2)$ and the integrals over the distribution amplitudes $\Phi_{\pm}(u; x^2)$.  
Thereby, it is necessary to have in mind that some of the lower moments of ${\varphi}_{A1|n}$ and ${\varphi}_{A2|n}$ vanish, related to corresponding Burkhardt-Cottingham sum rules.  

Let us first remark that the extra terms ${\varphi}_{A2|n}^{(4)}$ and ${\varphi}_{A2|n}^{(6)}$ contribute only to $\Phi_{-}(vx,x^2)$ thereby indicating that they are higher twist contributions as will be natural for the subleading $\Phi_{-}$. Furthermore, it is seen that by the additional ${\varphi}_{A2|n}$--terms the light-cone DAs 
$\Phi_{\pm}(v\lcx)$ are overdetermined. This was the reason to reject these terms in our previous work \cite{Bmeson}. Concerning the $x^2$--terms we observe that they are given in terms of differences ${\varphi}_{A1|n}^{2j} - {\varphi}_{A1|n}^{2(j+1)}$ and ${\varphi}_{A2|n}^{2j} - {\varphi}_{A2|n}^{2(j+1)}$ which seems to be the case also for higher orders of $x^2$.
Again, these additional terms are overdetermined if the ${\varphi}_{A2|n}$'s are taken into account, but would remain underdetermined and thus would lead to a difficult interplay between $\Phi_{+}(v\lcx)$ and $\Phi_{-}(v\lcx)$ if the ${\varphi}_{A2|n}$'s are absent. In the following subsection we demonstrate that these terms necessarily occur.
\medskip

\noindent
{\it (2) Derivation of Kawamura's et al. result from our decomposition into DAs of definite geometric twist}
\\
In Ref.~\cite{Kodaira2001} the $x^2$-dependence of $\Phi_{\pm|n}$ has not been made explicit and implicitly taken into consideration only partly. Therefore, let us take into account only the $x^2$-independent part of $\Phi_{\pm|n}$ by 
solving the local relations (\ref{R2}), (\ref{R1}) and (\ref{R4}) -- when the three-particle DAs are replaced by those of Ref.~\cite{Kodaira2001} -- in that order with respect to the independent variables 
${\varphi}_{A2|n}^{(4)}, {\varphi}_{A1|n}^{(4)}$ and ${\varphi}_{A1|n}^{(2)}$ for $n\geq 2$ as follows: 
\begin{align}
M^2\,{\varphi}_{A2|n}^{(4)} &= - n \int_0^1\!\!\d\vartheta\;\vartheta 
\left[ \Psi_{A|{n-2}} + X_{A|{n-2}} + 2\Psi_{V|{n-2}}\right]\!(\vartheta)\,,
\label{r2x}\\
M^2\,{\varphi}_{A1|n}^{(4)} &= 2 \int_0^1\!\!\d\vartheta\;\vartheta \Big\{
 (n-1)\left[ \Psi_{A|{n-2}} - \Psi_{V|{n-2}}\right]\!(\vartheta)
+(n+1)\left[ \Psi_{A|{n-2}} + X_{A|{n-2}} + 2\Psi_{V|{n-2}}\right]\!(\vartheta)
\Big\},
\label{r1x}\\
\intertext{}
M^2\,{\varphi}_{A1|n}^{(2)} &= \frac{2}{n+2} \Big\{\bar\Lambda M (n+1){\varphi}_{A1|n-1}^{(2)} 
- n(n-1) \int_0^1\!\!\d\vartheta\;\vartheta \left[ \Psi_{A|{n-2}} - \Psi_{V|{n-2}}\right]\!(\vartheta)
\nonumber\\
&\quad -(n+1)(n-1) \int_0^1\!\!\d\vartheta \left[\Psi_{A|{n-2}} + X_{A|{n-2}} + 2\vartheta \Psi_{V|{n-2}}\right]\!(\vartheta)
\Big\}.
\label{r4x}
\end{align}
Relation (\ref{R5}) could be resolved with respect to ${\varphi}_{A1|n}^{(6)}+ 2{\varphi}_{A2|n}^{(6)}$
using these expressions. But, taking the combination $(n-1)\,(\ref{R4}) - n \,(\ref{R5})$, 
we can avoid the use of expression (\ref{R4}). The result is (for $n\geq 3)$:
\begin{align} 
{\varphi}_{A1|n}^{(6)} +2 {\varphi}_{A2|n}^{(6)}
&= \frac{2 \,(n-1)}{M^2(n-2)}\int_0^1\!\!\d\vartheta\;(\vartheta -1) 
\Big\{(n-1)\left[ \Psi_{A|{n-2}} + X_{A|{n-2}}\right]\!(\vartheta)
			+ 2n \left[ \Psi_{A|{n-2}} + Y_{A|{n-2}}\right]\!(\vartheta)\Big\}
			\nonumber\\
&\quad 
	- \,\frac{2 \,(n-1)}{M^2(n-2)}\int_0^1\!\!\d\vartheta\;\vartheta \,
\Big\{\! \left[\Psi_{A|{n-2}} - \Psi_{V|{n-2}}\right]\!(\vartheta)
  +2 (n-2) \frac{\bar\Lambda}{M}
 \left[\Psi_{A|{n-3}} - \Psi_{V|{n-3}}\right]\!(\vartheta)\Big\}
	\nonumber\\
&\quad 
	- \, \frac{2}{M^2(n-2)}\int_0^1\!\!\d\vartheta\;\vartheta 
	\left[ \Psi_{A|{n-2}} + X_{A|{n-2}} + 2\Psi_{V|{n-2}}\right]\!(\vartheta)\,.
\label{r6x}	
\end{align}
From relation (\ref{ml}) we observe, 
that these twist-6 DAs occur in the $x^2$-dependent part of $\Phi_{-|n}$ only, but in the combination ${\varphi}_{A1|n}^{(6)}- 2{\varphi}_{A2|n}^{(6)}$. Therefore, a further relation seems to be missing. Possibly they 
may appear if the full $x^2$-dependence of all the two- and three-particle DAs is taken into account.

For $n=2$ the three-particle DAs are independent of $\vartheta$ and simply given by the double Mellin moments 
\begin{align}
\Psi_{A|0,0} = \hbox{\large$\frac{1}{3}$} \lambda_E^2, \qquad
\Psi_{V|0,0} = \hbox{\large$\frac{1}{3}$} \lambda_H^2, \qquad
X_{A|0,0} =  0, \qquad
Y_{A|0,0} =  0,
\end{align}
where $\lambda_E^2$ and $\lambda_H^2$ are given by the chromoelectric and chromomagnetic fields in the B-meson rest 
frame \cite{Grozin1996}.
Furthermore, according to (\ref{Y1b}) for $n=0$, as well as (\ref{R1}) and (\ref{R4}) for $n=1$ the following holds:
\begin{align}
\varphi_{A2|0}^{(4)} = 0,\qquad\qquad
\varphi_{A1|1}^{(4)} + 4 \varphi_{A2|1}^{(4)} = 0, \qquad\qquad
3 M \varphi_{A1|1}^{(2)} - 4 \bar\Lambda \varphi_{A1|0}^{(2)} =0.
\end{align}

Now, observing the normalization of $\Phi_{\pm|0}=1 $ let us determine the lowest moments of $ \Phi_{\pm|n}$
from the expressions (\ref{pl}) and (\ref{ml}):
\begin{align}
 \Phi_{+|0} &= -\varphi_{A1|0}^{(2)}=1,&	 &\Phi_{-|0} = -\varphi_{A1|0}^{(2)}=1,
 \\
 \Phi_{+|1} &= -\varphi_{A1|1}^{(2)}= \hbox{\large$\frac{4}{3}$} {\bar\Lambda}/{M},&	 
 &\Phi_{-|1} = -\f12\varphi_{A1|1}^{(2)}=\hbox{\large$\frac{2}{3}$} {\bar\Lambda}/{M},
 \\
 \Phi_{+|2} &= -\varphi_{A1|2}^{(2)}= 2 {\bar\Lambda}^2/{M^2} + \hbox{\large$\frac{2}{3}$}\lambda_E^2 + \hbox{\large$\frac{1}{3}$} \lambda_H^2,&	 
 &\Phi_{-|2} = -\hbox{\large$\frac{1}{3}$} \varphi_{A1|2}^{(2)} - \hbox{\large$\frac{2}{3}$} \varphi_{A1|2}^{(4)} 
 -2 \varphi_{A2|2}^{(4)}
 =\hbox{\large$\frac{2}{3}$}{\bar\Lambda}^2/{M^2} + \hbox{\large$\frac{1}{3}$} \lambda_H^2.
\end{align}
This is in full coincidence with the well-known result of \cite{Grozin1996} as well as Eqs.~(33) and (34) of Ref.~\cite{Kodaira2001}. 

Let us now consider the higher moments of $\Phi_{\pm|n}$. First, $\Phi_{+|n}$ has to be determined by solving  
 Eq. (\ref{R4}) iteratively with the result
\begin{align}
\Phi_{+|n} &= -{\varphi}_{A1|n}^{(2)}
=\frac{2}{n+2}\bigg\{\!\! \left( \frac{\bar\Lambda}{M}\right)^{\!\!n}
	+\; \sum_{k=0}^{n-2}\left( \frac{\bar\Lambda}{M}\right)^{\!\!k\;}
	\sum_{\ell=0}^{\,n-2-k}\binom{n-1-k}{\ell+1}\times 
	\nonumber\\
&\quad 
\bigg[\Big((n-k)\frac{2\ell +3}{\ell +2}+1\Big)\Psi_{A|n-2-k,\ell}
 + (n+1-k)X_{A|n-2-k,\ell} + (n+2-k) \frac{\ell+1}{\ell+2}\Psi_{V|n-2-k,\ell}\bigg]\bigg\},
\label{PHI+n}
\end{align}
where we used definition (\ref{DoubleMoments}) and performed the $\vartheta$-integrals. The first term of that expression, not containing  contributions from the three-particle DAs, by convention is called the Wandzura-Wilczek part $\Phi_{+|n}^{WW}$. 
(The notion ``Wandzura-Wilczek part'' is used quite differently in the Literature. Originally \cite{Wandzura1977} it was introduced to denote that contribution to the twist-3 structure function $g_2$ which appeared as ``geometric'' combination $g^{WW}_2(x) = -g_1(x) + \int_x^1 dy \,g_1(y)/y$ of the twist-2 structure function $g_1$, like the combination $\varphi_{A1}^{(2)}(u)- \varphi_{A1}^{(4)}(u)$ in Eq.~(\ref{pmnl}) appears when ignoring $2\,\varphi_{A2}^{(4)}(u)$; for a more detailed discussion of distinguishing between ``geometric'' and ``dynamic'' WW contributions, see e.g.~Refs.~\cite{BL01,Geyer2000,Lazar2000}).
Remark also that there appears no term $\left( {\bar\Lambda}/{M}\right)^{n-1}$ because the three-particle DAs start with $n=2$. 

Next, we express $\Phi_{-|n}$ by $\Phi_{+|n}$ plus an extra term as follows
\begin{align}
\Phi_{-|n} 
-\frac{1}{n+1}\,\Phi_{+|n}
=-\left[\frac{n}{n+1}\,{\varphi}_{A1|n}^{(4)}+2{\varphi}_{A2|n}^{(4)}\right]
= -\frac{2n}{n+1}\sum_{\ell=0}^{n-2}\binom{n-1}{\ell+1}\frac{\ell+1}{\ell+2}\left[\Psi_A -\Psi_V\right]_{n-2,\ell}.
\label{PHI-n}
\end{align}
Up to a trivial change of indexes expressions (\ref{PHI+n}) and (\ref{PHI-n}) coincide with the final result of Kawamura et al. given by formulas (28) -- (31) of Ref.~\cite{Kodaira2001}.

From this result we observe that\\
$\bullet$~~despite not exhausting all informations contained in the EoM the result of Ref.~\cite{Kodaira2001} is complete as far as the DAs $\Phi_{\pm|n}$ {\em on the light-cone} are concerned, \\ 
$\bullet$~~from the point of view of geometric twist $\Phi_{+|n}$ really is of minimal twist $\tau = 2$, but $\Phi_{-|n}$
contains also a (smaller) part of minimal twist $\tau = 2$ together with two different parts of higher twist $\tau = 4$,
\\
$\bullet$~~the above derivation of Eqs. (\ref{PHI+n}) and (\ref{PHI-n}) shows the necessity of maintaining the contributions ${\varphi}_{A2|n}^{(\tau)}$ since otherwise these expressions would read quite different in contradistinction to Ref.~\cite{Kodaira2001} and, furthermore, due to Eq.~(\ref{R2}) we were led to 
$\int_0^1\!\!\d\vartheta\;\vartheta \left[\Psi_{A|{n-2}} + X_{A|{n-2}} + 2\Psi_{V|{n-2}}\right]\!(\vartheta) = 0$ which together with (\ref{Z2}) and (\ref{Z4}) had strange consequences.
\medskip

\noindent
{\it (3) Some aspects concerning the $x^2$-dependence of $\Phi_{\pm}(vx, x^2)$ resp. transverse momentum dependence of their Mellin moments}
\\
Looking at (\ref{pl}) and (\ref{ml}) we find that $\Phi_{\pm}(vx, x^2)$ for $n\geq 2$ contain $x^2$-dependent terms which are related to the differences ${\varphi}_{A1|n}^{(2)} - {\varphi}_{A1|n}^{(4)}$ and 
${\varphi}_{Ai|n}^{(4)} - {\varphi}_{Ai|n}^{(6)}, i = 1,2$ of DAs of consecutive twists. Of course, these $x^2$-dependent terms for the Mellin moments lead to contributions of transverse momenta $\vec{k}_\bot$. 

In Refs.~\cite{Kodaira2003,HWZ05} omitting the contribution of 3-particle DAs, i.e.~in ``Wandzura-Wilczek-Approximation'',
the corresponding $x^2$-dependent equations for the Fourier transform with respect to the longitudinal separation $t= vx$ (for convenience, we change to the variable $\omega = M\,u$),
\begin{align}
 \Phi_\pm^{WW}(\omega, x^2) = \int \frac{\d t}{2\pi}\; e^{\,\ii \,\omega \,t}\,\Phi_\pm^{WW}(t, x^2)\,,
\end{align}
has been exactly solved whereby they found that both functions have a common $x^2$-dependence:
\begin{align}
 \Phi_\pm^{WW}(\omega, x^2) &= \phi_\pm^{WW}(\omega) \, \chi\left(x^2\, \omega( 2\bar\Lambda -\omega ) \right)\,,
 \label{WW1}
\\
\mathrm{where} \qquad \qquad\qquad \phi_\pm^{WW}(\omega) &= \frac{\bar\Lambda\pm(\omega-\bar\Lambda)}{2\bar\Lambda^2}\,\theta(\omega)\theta(2\bar\Lambda - \omega), 
\\
\mathrm{and} \qquad\quad\;
\chi\left(x^2\,\omega ( 2\bar\Lambda - \omega) \right)&= J_0\left(|\mathbf{x}_\bot|\sqrt{\omega ( 2\bar\Lambda - \omega)}\,\right)
 \quad \mathrm{with}\quad x^2 = - \mathbf{x}^2_\bot \,.
  \label{WW3}
\end{align}
In principle, we would be in a position to confirm that result since we know from expressions (\ref{Y1}) and (\ref{Y1hat}) the full $x^2$-dependence of $\Phi_\pm^{WW}(t, x^2)$ but had to perform the Fourier transformation which is by no means simple. However, since the solution (\ref{WW1}) -- (\ref{WW3}) is an exact one we stop here.

In Ref.~\cite{HQW2006} the full problem has been tackled by assuming a common transverse momentum dependence also for each of the (Fourier transformed) 3-particle DAs. Despite being plausible that assumption has to be verified.
In principle, the $x^2$-dependence of the three-particle DAs can be determined in the same manner as we did it for the two-particle DAs by using the projections onto operators of definite geometric twist as given in the Appendix.
This remains an open problem which must be postponed to another paper.

In addition, some simplifying relations between the 3-particle DAs are introduced. And finally, a special model for the difference $\Psi_A(u_1,u_2) - \Psi_V(u_1,u_2)$ is required which, together with the other two requirements leads to the restriction
$ Y_A(u_1,u_2) = -\,\Psi_A(u_1,u_2) = X_A(u_1,u_2)$ 
and, from the view of the present work, seems to be very stringent.

\setcounter{equation}{0}
\section{Summary and Concluding Remarks}
\label{conclusion}
\label{EoMconcl} 

With the aim of extending our previous work \cite{Bmeson} we used our knowledge about the explicit off-cone structure of QCD tensor operators of definite geometric twist (up to tensors of second stage) for rewriting the (relevant) EoM -- connecting the heavy mesons two- and three-particle DAs on the light-cone -- into a set of algebraic equations for the (double) Mellin moments corresponding to these amplitudes. Thereby we have taken into account the heavy quark on-shell constraint and the well-known relations between Dirac's $\gamma$-matrices in order to show that, in principle, it is sufficient to restrict to the axial vector structure $\Gamma=\gamma_5\gamma_\alpha$.

First, we found that two types of two-particle distribution amplitudes, $\varphi_{A1}^{(\tau)}$ [for $\tau = 2,4,6$] and $\varphi_{A2}^{(\tau)}$ [for $\tau = 4,6$], and five types of three-particle distribution amplitudes, $\Upsilon_{T1}^{(4)}, \Upsilon_{T2}^{(\tau)}$ [for $\tau = 4,6$] and $\Upsilon_{T3}^{(6)}$ as well as $\Omega_{A1}^{(\tau)}$ [for $\tau= 3,5$] and $\Omega_{A2}^{(5)}$ occur; higher twists would appear if the EoM were considered off the light-cone. In comparison with our previous work we introduced two additional types of DAs.
These sets of independent two- and three-particle DAs of definite twist are much larger than the commonly used sets of DAs consisting of $\Phi_\pm$ as well as $\Psi_V, \Psi_A, X_A$ and $Y_A$, respectively. In Sect.~\ref{EoMdisc} by comparing the representation of the various matrix elements in terms of the corresponding DAs we derived the relations between conventional DAs and and those of definite geometric twist, Eqs.~(\ref{K1}) -- (\ref{K4}), together with three relations, Eqs.~(\ref{Z01}) -- (\ref{Z03}), connecting part of the DAs of definite geometric twist -- some of them as special combinations -- with two further sets, $\Upsilon_{V}^{(4)}$ and $\Upsilon_{P}^{(5)}$. Thereby, we were able to select four appropriate combinations of three-particle DAs of definite twist, Eqs.~(\ref{Th1n}) -- (\ref{Th4n}), which can be used to simplify the representation of the four EoM under study. In Sect.~\ref{EoMKodaira} the conventional two-particle DAs $\Phi_\pm$ are represented (up to first order in $x^2$) by the above mentioned DAs $\varphi_{Ai}^{(\tau)}, i=1,2$, Eqs.~(\ref{pl}) -- (\ref{mnl}), showing that already on the light-cone the conventional ones contain three independent DAs of definite twist.

Next, in Sect.~\ref{EoMrelations}, we presented four sets of algebraic equations, Eqs.~(\ref{R1}) and (\ref{R2}) as well as (\ref{R4}) and (\ref{R5}), connecting some combinations of the two-particle Mellin moments of definite twist, $\varphi_{Ai|n}^{(\tau)}, i = 1,2$, with the $\vartheta$-integrated combinations (\ref{Th1n}) -- (\ref{Th4n}) of ($\vartheta$-dependent sums of) three-particle double Mellin moments of definite twist. In addition, two independent $\vartheta$-integrated relations, Eqs.~(\ref{Z2}) and (\ref{Z4}), between the (double) Mellin moments are derived by combining the two kinds of EoM, Eqs.~(\ref{X4a}) and (\ref{X4}), with the relation (\ref{X2}) which was due to the Chisholm identity. In Sect.~\ref{EoMnonlocalrelations} these relations are reformulated non-locally as relations (\ref{R1n}) -- (\ref{Z4nl}) between the two- and three-particle DAs. Furthermore, the relations (\ref{Z01}) -- (\ref{Z03}) are reformulated non-locally in relations (\ref{KZ02nl}), (\ref{Omega5n0l}) and (\ref{R2n0l}). Also in Sect.~\ref{EoMnonlocalrelations} we presented the vanishing of lower (double) Mellin momenta, Eqs.~(\ref{Y1a}) and (\ref{Y1b}) as well as (\ref{Y01}) -- (\ref{Y04}), by corresponding Burkhardt-Cottingham-like sum rules Eqs.~(\ref{U1}) and (\ref{U2}) as well as (\ref{U3}) -- (\ref{U6}). Obviously, the results of Sects.~\ref{EoMrelations} and \ref{EoMnonlocalrelations} are the main results of this paper.
 
Finally, in Sect.~\ref{EoMKodaira}, by resolving Eqs.~(\ref{R1}), (\ref{R2}), (\ref{R4}) and (\ref{R5}) with respect to the independent two-particle Mellin moments ${\varphi}_{A2|n}^{(4)}, {\varphi}_{A1|n}^{(4)}, {\varphi}_{A1|n}^{(2)}$ and ${\varphi}_{A1|n}^{(6)}+ 2{\varphi}_{A2|n}^{(6)}$, Eqs.~(\ref{r2x}) -- (\ref{r6x}), we were able, as a consistence check, to re-derive the result of Kawamura et al.~\cite{Kodaira2001} for the Mellin moments $\Phi_{\pm|n}$, given here by Eqs.~(\ref{PHI+n}) and (\ref{PHI-n}). In principle, the relations (\ref{r2x}) -- (\ref{r6x}) can be used to express $\Phi_{\pm|n}(x^2)$ at least (partially) up to order $x^2$ by the three-particle double Mellin moments --- if not a further relation were missing which would allow to separate ${\varphi}_{A1|n}^{(6)}$ and ${\varphi}_{A2|n}^{(6)}$. To our opinion this requires the consideration of the $x^2$-dependence in next order of all the participating two- and three-particle DAs of definite twist which, however, was not the aim of the present paper.
For that reason we also were not able to further comment on the transverse momentum dependence of the two-particle DAs and their relation to the three-particle DAs.

Despite being mathematically more complex, the group theoretically motivated use of the notion of geometric twist allows for a very clear distinction between the contributions of different twist to the various matrix elements of physically relevant QCD operators and the corresponding DAs. Based on the quantum field theoretical framework, see Refs.~\cite{Zavialov,Geyer1994}, it allows also for a different look at the conventionally introduced DAs, e.g., concerning the support of the various DAs or the appearance of new sum rules. 
Our study also showed that the twist structure of $\Phi_{-|n}$ is more complicated than usually assumed and requires an additional DA of twist-4; in addition, the usual three-particle DAs are shown to be appropriate combinations of three-particle DAs of definite geometric twist, thereby $Y_{A|n}$ and $\Psi_{A|n}$ appear as complicated combinations of three different DAs each, compare expressions (\ref{T1}) -- (\ref{T4}) with (\ref{Th1}) -- (\ref{Th4}), cf. also relations (\ref{K1}) -- (\ref{Z03}). These results bring some light onto the recently raised question \cite{KMO} if the three-particle matrix element (\ref{Kodaira_tri0}) requires the introduction of additional DAs.

Furthermore, from the point of view of the present paper, it seems to be possible to solve in that manner the problem of determining the transverse momentum dependence of the heavy mesons wave functions. Of course, concerning the two-particle DAs this requires the consequent use of Eqs.~(\ref{Y1}) and (\ref{Y1hat}) and, additionally, to find the (infinite) off-cone decomposition of Eqs.~(\ref{Dtri_v}) -- (\ref{F1}) and (\ref{Dtri_vv}) -- (\ref{Dtri_g_v}); the latter becomes somewhat more difficult because of the appearance of additional independent twist structures.

\noindent
{\large\bf Acknowledgement}\\
The authors are very much indebted to J\"org Eilers for informing them about his results on the off-cone twist decomposition of QCD tensor operators up to second stage before preliminary publication (arXiv: hep-th/0608173)
as well as many discussions on that and related matter. OW acknowledges support by the DFG within the SFB/TR 9 ``Computational Particle Physics''.


 \renewcommand{\theequation}{\thesection.\arabic{equation}}\setcounter{equation}{0}
\begin{appendix}
\setcounter{equation}{0}
\section{Off-cone tensor operators of definite geometric twist}
\label{Append1}
Actually, we are interested in obtaining the twist decomposition of bilocal as well 
as of trilocal operators. As mentioned, only the tensorial structure of the operator 
is crucial for calculation and not whether it is created bilocally or  trilocally. 
Hence, we deal with trilocal operators as if they were bilocal, but have to take care 
of the third field when writing the results. 

We denote the generic nonlocal off-cone (pseudo) scalar, (axial) vector and 
second rank tensor operators as follows:
\begin{align}
N(\kappa_1 x,\kappa_2 x),
\qquad 
O_\alpha(\kappa_1 x,\kappa_2 x),
\qquad 
M_{\alpha\beta}(\kappa_1 x,\kappa_2 x),
\nonumber
\end{align}
where a possible pseudo structure is not labeled. The tensor operator of second rank 
splits up in an antisymmetric part $M_{[\alpha\beta]}(\kappa_1 x,\kappa_2 x)$ and a 
symmetric part $M_{(\alpha\beta)}(\kappa_1 x,\kappa_2 x)$ from which also the trace 
$M(\kappa_1 x,\kappa_2 x)={M_{\alpha}}^{\alpha}(\kappa_1 x,\kappa_2 x)$ could be taken. 
The corresponding light-cone operators are obtained by replacing
$x\to \lcx$ and applying the constraint $\lcx^2=0$.

In principle, we can perform the twist decomposition equivalently in the nonlocal representation as well as in the local representation. Here, we choose the local one. 
The relations between the nonlocal and the local operators are given by the Taylor
expansion according to (restricting to $\kappa_1 =\kappa;\;\kappa_2=0$):
\begin{align}
N(\kappa x,0) = \sum_{n=0}^\infty \frac{\kappa^n}{n!} N_n(x),
\qquad
O_\alpha(\kappa x,0) = \sum_{n=0}^\infty \frac{\kappa^n}{n!} O_{\alpha| n}(x),
\qquad
M_{\alpha\beta}(\kappa x,0) = \sum_{n=0}^\infty \frac{\kappa^n}{n!} M_{\alpha\beta| n}(x).
\label{Taylor}
\end{align}

For the sake of a compact notation and also in order to make obvious the relation 
between the on-cone and off-cone version, we introduce the `interior' differential 
operator which on the light-cone is given by \cite{Bargmann1977}:
\begin{align}
\lcdi_\alpha f(\lcx)
&= \left\{(1+x\partial) \partial_\alpha -\hbox{$\frac{1}{2}$} x_\alpha 
\Box\right\}\! f(x) \big|_{\;x=\lcx}\,.
\label{lind}
\end{align} 
Its harmonic off-cone extension $\di$ and the complementary off-cone $\ix$-operator 
are given as follows:
\begin{align}
\di_\alpha = (1+x\partial) \partial_\alpha -\hbox{$\frac{1}{2}$} x_\alpha \Box,
\qquad
\ix_\alpha = x_\alpha (1+x\partial) - \hbox{$\frac{1}{2}$} x^2 \partial_\alpha.
\label{ind}
\end{align}
The operators $\di_\alpha, x_\alpha, X = 1 + x\pd$ and 
$X_{\alpha\beta}= x_\beta \pd_\alpha - x_\alpha \pd_\beta$ 
span the conformal algebra $so(4,2)$ as do the corresponding `interior' operators
on the light-cone, cf.~Refs.~\cite{Bargmann1977},\cite{Joerg}. Especially, the off-cone
operators obey also the following relations
\begin{gather}
(X-1)\,\di_{[\alpha} x_{\beta]} = - \,(X+1)\, x_{[\alpha} \di_{\beta]}\,,
\qquad
\label{ac}
\di_{(\alpha} x_{\beta)} =  x_{(\alpha} \di_{\beta)} +  X g_{\alpha\beta}\,,
\end{gather} 

Off the light-cone, the decomposition of a tensor operator of finite rank $r$ with 
respect to the irreducible representations of the orthochronous Lorentz group is
given by a finite series of traceless tensors having a well-defined symmetry type,
cf.,~e.g.~\cite{Barut1977}. In our case, the possible symmetry types are restricted 
by the fact that any generic local operator ${\cal O}_{\Gamma|n}(x)$ resulting from
a Taylor expansion is a homogeneous polynomial of order $n$ and completely symmetric 
w.r.t.~the $n$ indices which are truncated by the vectors $x$. However, this also
allows the application of the polynomial technique \cite{Bargmann1977}. Consequently,
these polynomials will vanish if more than $n$ derivatives act on it. Therefore, 
the projectors exhibit an intrinsic termination which avoids the occurrence of undefined 
fractions or factorials (see below) \cite{Eilers2003}. 

The twist decomposition of the relevant local tensor operators is given by 
Eqs.~(\ref{TWSg}) -- (\ref{TWaTu}) below. Thereby, 
the various contributions of twist $\tau=$ (canonical) dimension $-$ (Lorentz) spin 
are labeled by $\tau_0$ plus the higher order contribution due to the decomposition with 
respect to the irreducible representations of $SO(3,1)$. Thereby, $\tau_0$ is defined as 
the twist corresponding to the (fictitious) entirely symmetrized operator; in cases, 
where the operator can exhibit entire symmetry, $\tau_0$ also denotes the minimal twist 
of that operator and in cases, where entire symmetry is not allowed for that operator, 
as for the antisymmetric tensor of second rank, we use $\tau_0$ as a counter only, cf.
\cite{Eilers2003}. The contributions resulting from subtractions of traces are numbered
by $j$ which count powers of $x^2$ (accompanied by powers of $\Box$) in the projector. 

All off-cone projection operators include the following projectors $H_n$ onto traceless 
homogeneous polynomials of degree $n$ \cite{Barut1977}, 
\begin{align}
H_n (x^2;\Box)
&=\sum_{k=0}^{[\frac{n}{2}]} \frac{(-1)^k (n-k)!}{4^k\, k!\, n!} (x^2)^k \Box^k \,,
\end{align}
which are sufficient for the formulation of the off-cone twist decomposition of scalar
operators.
For the vector and tensor case they contain, in addition, also specific tensor operators
which are related to the symmetry type of the different twist contributions. 
The scalar projection operator has contributions related to even spin only 
whereas vector and skew-tensor projection operators exhibit even and odd contributions:
\begin{align}
N^{(\tau_0+2j)}_n(x) 
&=  \frac{(n+1-2j)!}{4^j\, j!\,(n+1-j)!}\; (x^2)^j\, 
H_{n-2j}(x^2|\Box)\;\Box^j \,N_n(x); 
 \label{TWSg}
 \\
 \nonumber \\
O^{(\tau_0+2j)}_{\alpha|n}(x) 
&= \frac{1}{4^j\,j!}\;
\bigg\{\frac{(n+1-2j)!}{(n+1-j)!}\;(x^2)^{j}\;
\frac{H_{n-2j}(x^2|\Box)\;\di_\alpha x^\mu}{(n+1-2j)^2}\;\Box^j 
\nonumber \\
&\quad 
+ 4j\,
\frac{\big(n+3-2j\big)!}{\big(n+2-j\big)!}\;(x^2)^{j-1}\;
\frac{H_{n+2-2j}(x^2|\Box)\;x_\alpha \di^\mu}{(n+3-2j)^2}\;\Box^{j-1} 
\bigg\}O_{\mu|n}(x),
\label{TWVg}
\\
O^{(\tau_0+1+2j)}_{\alpha|n}(x) 
&=  \frac{(n+1-2j)!}{4^j\, j!\,(n+1-j)!}\, 
(x^2)^j \;H_{n-2j}(x^2|\Box)\bigg[
\delta_\alpha^\mu
-\frac{x_\alpha \di^\mu + \di_\alpha x^\mu }{(n+1-2j)^2}
\bigg] \Box^j\; O_{\mu|n}(x);
\label{TWVu}
\\
\nonumber\\
M^{(\tau_0+1+2j)}_{[\alpha\beta]|n}(x)
&= \frac{-\,2}{4^j\, j!} 
\bigg\{\frac{(n+1-2j)!}{(n+1-j)!}\,
\frac{(x^2)^{j}\,H_{n-2j}(x^2|\Box)}{(n+1-2j)(n+2-2j)}\,
\bigg[
\di_{[\alpha}^{\phantom{[\mu}}\delta_{\beta]}^{[\mu}\,x^{\nu]}_{\phantom{\beta]}}
-\frac{x_{[\alpha} \di_{\beta]}\,x^{[\mu}\di^{\nu]} }{(n-2j)^2}\;
\bigg]\;\Box^j
\label{TWaTg}
\\
&\quad
+4j\,\frac{(n+3-2j)!}{(n+2-j)!}\,
\frac{(x^2)^{j-1}\,H_{n+2-2j}(x^2|\Box)}{(n+2-2j)(n+3-2j)} 
\bigg[
x_{[\alpha}^{\phantom{[\mu}}\delta_{\beta]}^{[\mu}\,\di^{\nu]}_{\phantom{\beta]}}
-\frac{x_{[\alpha} \di_{\beta]}\,x^{[\mu}\di^{\nu]}}{(n+2-2j)^2}
\bigg] \Box^{j-1} \!
\bigg\} M_{[\mu\nu]|n}(x),
\nonumber
\\
M^{(\tau_0+2+2j)}_{[\alpha\beta]|n}(x)
&= \frac{1}{4^j\, j!}
\bigg\{ \frac{(n+1-2j)!}{(n+1-j)!} 
(x^2)^j\, H_{n-2j}(x^2|\Box)
\bigg[
\delta_{[\alpha}^{[\mu} \delta_{\beta]}^{\nu]} 
\nonumber \\
&\quad
+\frac{2}{n+1-2j}
\bigg(\,
 \frac{x_{[\alpha}^{\phantom{[\mu}}\delta_{\beta]}^{[\mu}\, \di^{\nu]}_{\phantom{\beta]}}}{n-2j}\;
+\frac{\di_{[\alpha}^{\phantom{[\mu}}\delta_{\beta]}^{[\mu}\, 
x^{\nu]}_{\phantom{\beta]}} }{n+2-2j}\,
\bigg)
-\frac{2\,x_{[\alpha} \di_{\beta]}\, x^{[\mu} \di^{\nu]} }{(n-2j)^3(n+2-2j)}
\bigg] \Box^j 
\label{TWaTu}
\\
& \quad
- 4j\, \frac{(n+3-2j)!}{(n+2-j)!} \,(x^2)^{j-1}\,H_{n+2-2j}(x^2|\Box)
\frac{2\,x_{[\alpha} \di_{\beta]}\,x^{[\mu}\di^{\nu]} }{(n+2-2j)^3(n+4-2j)}
\,\Box^{j-1} \!\bigg\} 
M_{[\mu\nu]|n}(x) .
\nonumber
\\
\nonumber \\
M^{(\tau_0+2j)}_{(\alpha\beta)|n}(x) 
&= \frac{1}{4^j\,j!}\bigg\{
\frac{(n+1-2j)!}{(n+1-j)!}\;
\frac{(x^2)^{j}\;H_{n-2j}(x^2|\Box)}{(n+1-2j)^2(n+2-2j)^2}\;
\di_{\alpha} \di_{\beta} x^{\mu} x^{\nu}\Box^j
\nonumber \\
& \quad 
+4j\,\frac{(n+3-2j)!}{(n+2-j)!}\;
(x^2)^{j-1}\;H_{n+2-2j}(x^2|\Box)
\bigg[\frac{1}{2} \delta_{\alpha\beta} \delta^{\mu\nu}
\nonumber \\
& \qquad \qquad 
-  \frac{x_{(\alpha} \di_{\beta)} \delta^{\mu\nu}
      + \delta_{\alpha\beta} \di^{(\mu} x^{\nu)}}{(n+4-2j)^2}
+  \frac{4 x_{(\alpha} \di_{\beta)} x^{(\mu} \di^{\nu)}}{(n+2-2j)^2(n+4-2j)^2}
\bigg]\Box^{j-1}
\label{TWsTg1}
\\
& \quad 
+16j(j-1)\frac{(n+5-2j)!}{(n+3-j)!}\;
\frac{(x^2)^{j-2}\;H_{n+4-2j}(x^2|\Box)}{(n+4-2j)^2(n+5-2j)^2}
x_{\alpha} x_{\beta} \di^{\mu} \di^{\nu} \Box^{j-2}
\bigg\}M_{(\mu\nu)|n}(x)\,, 
\nonumber
\\
M^{(\tau_0+1+2j)}_{(\alpha\beta)|n}(x) 
&= \frac{2}{4^j\,j!}\;
\bigg\{\frac{(n+1-2j)!}{(n+1-j)!}\;
\frac{(x^2)^{j}\;H_{n-2j}(x^2|\Box)}{(n-2j)(n+1-2j)}
\di_{(\alpha}\bigg[ 
\delta_{\beta)}^{(\mu} 
-
\frac{x_{\beta)} \di^{(\mu} + \di_{\beta)} x^{(\mu}}{(n+2-2j)^2}
\bigg]x^{\nu)} \Box^j
\label{TWsTu}
\\
&~
+ 4j\,
\frac{\big(n+3-2j\big)!}{\big(n+2-j\big)!}\;
\frac{(x^2)^{j-1}\;H_{n+2-2j}(x^2|\Box)}{(n+3-2j)(n+4-2j)}
x_{(\alpha}\bigg[\delta_{\beta)}^{(\mu}
- \frac{x_{\beta)}\di^{(\mu}+\di_{\beta)} x^{(\mu}}{(n+2-2j)^2}
\bigg]\di^{\nu)}\Box^{j-1}\!
\bigg\} M_{(\mu\nu)|n}(x),
\nonumber\\
\intertext{}
M^{(\tau_0+2+2j)}_{(\alpha\beta)|n}(x) 
&= 
\frac{(n+1-2j)!}{4^j\,j!\,(n+1-j)!}\;
(x^2)^{j}\;H_{n-2j}(x^2|\Box)
\bigg[
\delta_{(\alpha}^{(\mu} \delta_{\beta)}^{\nu)}
- \frac{2}{n+1-2j}\Big(
 \frac{x_{(\alpha} \delta_{\beta)}^{(\mu}\di^{\nu)}}{n+2-2j} 
+\frac{\di_{(\alpha} \delta_{\beta)}^{(\mu}  x^{\nu)} }{n-2j} 
\Big)
\nonumber \\
&\quad 
+\frac{\di_{\alpha} \di_{\beta} x^{\mu} x^{\nu}
+ x_{\alpha} x_{\beta} \di^{\mu} \di^{\nu}}{(n-2j)(n+1-2j)^2(n+2-2j)}
+\frac{x_{(\alpha} \di_{\beta)} \delta^{\mu\nu}
+ \delta_{\alpha\beta} \di^{(\mu} x^{\nu)}}{(n-2j)(n+2-2j)}
- \frac{1}{2}\delta_{\alpha\beta} \delta^{\mu\nu}
\bigg]\Box^j M_{(\mu\nu)|n}(x)\,. 
\label{TWsTg2}
\end{align}

The finite series of scalar off-cone operators of definite twist (\ref{TWSg}) have 
been given in Ref.~\cite{Geyer1999} using the harmonic extension of corresponding 
light-cone functions \cite{Bargmann1977}. 
The off-cone vector operators of definite twist (\ref{TWVg}) and (\ref{TWVu}) have 
been determined for the first time in \cite{Eilers2003} but are reformulated here
appropriately; the expressions of even twist $\tau_0+2j$ contain two different series 
of operators of definite twist with the second series starting at $j=1$.
The off-cone antisymmetric tensor operators of definite twist (\ref{TWaTg}) and 
(\ref{TWaTu}) and the off-cone symmetric tensor operators of definite twist 
(\ref{TWsTg1}) -- (\ref{TWsTg2}) in any dimension have been given for the first 
time in Ref.~\cite{Joerg}; they are reformulated here appropriately for 
$D \equiv 2h = 4$ dimensions. In the antisymmetric case the contributions of
even and odd twist, $\tau_0+1+2j$ and $\tau_0+2+2j$, contain two different series of 
operators of definite twist where, again, the second ones begin with $j=1$.
In the symmetric case the contributions of even twist, $\tau_0+2j$ and $\tau_0+2+2j$, 
contain three different and a single series of operators of definite twist, respectively, 
and the contributions of odd twist, $\tau_0+1+2j$, contain two different series of 
operators of definite twist.

All the series of local operators of definite twist terminate at 
$j_{\mathrm max} = \big[\frac{n+f}{2}\big]$ with $f$ being the number of free 
tensor indices $\Gamma$. Summing up all the twist contributions results in the non-decomposed 
local operator showing that the twist decomposition is a decomposition of unity into
traceless tensors of definite symmetry type, i.e. irreducible representations of the
Lorentz group.
Obviously, the twist projection operators are applied to the non-decomposed operators: 
\begin{align}
N^{(\tau)}_n(x) = \left({\cal P}^{(\tau)} N_{n}\right) (x),
\qquad 
O^{(\tau)}_{\alpha|n}(x) = \left({\cal P}_\alpha^{(\tau)\mu} O_{\mu|n}\right) (x),
\qquad 
M^{(\tau)}_{[\alpha\beta]|n}(x) = \left({\cal P}_{[\alpha\beta]}^{(\tau)[\mu\nu]} M_{[\mu\nu]|n}\right)(x).
\nonumber
\end{align}
They obey the well-known properties of projection operators,
\begin{align}
\big(\widetilde{\cal P}^{(\tau)} \times 
\widetilde{\cal P}^{(\tau')}\big)^{\Gamma' n'}_{\Gamma n}
= \delta^{\tau \tau'}\widetilde{\cal P}^{(\tau)\Gamma' n'}_{~\Gamma n},
\qquad 
\sum_{\tau = \tau_{\rm min}}^{\tau_{\rm max}}\widetilde{\cal P}^{(\tau)}
= \bf{1}.
\nonumber
\end{align}

Let us remark, that the reduction onto the light-cone obtains simply by using 
$\di \rightarrow \lcdi$ and $x \rightarrow \lcx,\;\lcx^2 = 0,$ resulting, especially, 
in $H_{n}(\lcx^2|\Box)=1$ and restriction to $j=0,1$. These on-cone operators have 
been given already in our previous work \cite{Bmeson}, Eqs.~(I.B.8) -- (I.B.19).
Thereby, besides Eqs.~(\ref{ac}), the following useful relations should be obeyed:
\begin{gather}
\ix_\alpha \,H_{n}(x^2|\Box)= H_{n+1}(x^2|\Box)\,x_\alpha (1+x\pd ),
\qquad
(1+x\pd )\pd_\alpha \,H_{n}(x^2|\Box)= H_{n-1}(x^2|\Box) \,\di_\alpha,
\nonumber\\
\ix_\alpha \pd_\beta \,H_{n}(x^2|\Box)= H_{n}(x^2|\Box) \,x_\alpha \di_\beta,
\qquad
\pd_\alpha \ix_\beta \,H_{n}(x^2|\Box)= H_{n}(x^2|\Box)\, \di_\alpha x_\beta.
\nonumber
\end{gather}

Here we presented the twist projectors thereby putting together different contributions of 
the same twist but of different symmetry type since we are interested in applying them for 
the twist decomposition of a distribution amplitudes which are related to their non-forward matrix elements in a given parametrization. From the group-theoretical point of view, such 
a summation losses some information about the structure of the distribution amplitudes which, however, could be re-covered, if necessary.  
For a more detailed discussion, see, Refs.~\cite{Eilers2003,Joerg}.

The local operators ${\cal O}^{(\tau)}_{\Gamma|n}(x)$ of definite twist can be summed
up into non-local operators ${\cal O}^{(\tau)}_{\Gamma}(\kappa x,0)$ of definite twist 
by appropriately rewriting the various fractions in $n$ which appear in Eqs.~(\ref{TWSg}) 
-- (\ref{TWaTu}) in terms of integrals thus arriving at the generic forms (\ref{Taylor}).
This also works in the case of distribution amplitudes when determining it from their
Mellin moments in Section \ref{EoMdisc}.

\end{appendix}

\newpage


\end{document}